\documentclass[journal]{IEEEtran} 
\ifCLASSINFOpdf
\else
\fi
\usepackage[cmex10]{amsmath}
\usepackage{array}
\usepackage{multirow}

\ifCLASSOPTIONcompsoc
  \usepackage[caption=false,font=normalsize,labelfont=sf,textfont=sf]{subfig}
\else
  \usepackage[caption=false,font=footnotesize]{subfig}
\fi
\pdfoutput=1
\usepackage{graphicx} 
\graphicspath{{figure/}}
\DeclareGraphicsExtensions{.pdf}
\usepackage{fancyhdr}
\usepackage{hyperref}

\begin{document}

%
\title{Perceptual experience analysis for tone-mapped HDR videos based on EEG and peripheral physiological signals}
%
%
%

\author{Seong-Eun~Moon,
        Jong-Seok~Lee,~\IEEEmembership{Senior Member,~IEEE,}
\thanks{The authors are with the School of Integrated Technology and Yonsei Institute of Convergence Technology, Yonsei University, Incheon 406-840, Republic of Korea
(e-mail: \{se.moon, jong-seok.lee\}@yonsei.ac.kr).}
\thanks{Manuscript received November 10, 2014; revised May 6, 2015.}
\thanks{\copyright 2015 IEEE. Personal use of this material is permitted. Permission from IEEE must be obtained for all other uses, in any current or future media, including reprinting/republishing this material for advertising or promotional purposes, creating new
collective works, for resale or redistribution to servers or lists, or reuse of any copyrighted
component of this work in other works. }
\thanks{This is the author's version of an article that has been published in this journal (DOI: 10.1109/TAMD.2015.2449553). Changes were made to this version by the publisher prior to publication. The final version of record is available at \url{http://dx.doi.org/10.1109/TAMD.2015.2449553}}
}

%
%

\markboth{IEEE transactions on autonomous mental development}%
{}
%



\maketitle

\begin{abstract}
High dynamic range (HDR) imaging has been attracting much attention as a technology that can provide immersive experience. Its ultimate goal is to provide better quality of experience (QoE) via enhanced contrast. In this paper, we analyze perceptual experience of tone-mapped HDR videos both explicitly by conducting a subjective questionnaire assessment and implicitly by using EEG and peripheral physiological signals. From the results of the subjective assessment, it is revealed that tone-mapped HDR videos are more interesting and more natural, and give better quality than low dynamic range (LDR) videos. Physiological signals were recorded during watching tone-mapped HDR and LDR videos, and classification systems are constructed to explore perceptual difference captured by the physiological signals. Significant difference in the physiological signals is observed between tone-mapped HDR and LDR videos in the classification under both a subject-dependent and a subject-independent scenarios. Also, significant difference in the signals between high versus low perceived contrast and overall quality is detected via classification under the subject-dependent scenario. Moreover, it is shown that features extracted from the gamma frequency band are effective for classification. 
\end{abstract}

\begin{IEEEkeywords}
High dynamic range video, electroencephalography (EEG), physiological signal, quality of experience (QoE).
\end{IEEEkeywords}

%
\IEEEpeerreviewmaketitle

\section{Introduction}
%
%
%
%
\IEEEPARstart{I}{n} recent years, demand for realistic and immersive multimedia contents is increasing. Following this trend, high dynamic range (HDR) imaging has been attracting much attention as one of the technologies that can meet such end users' demand. The dynamic range usually means the ratio between the maximum and minimum values that a sensor can measure. In photography, the dynamic range means the luminance range of a captured scene or the limit of luminance range that a given digital camera can capture. HDR image and video contents have extended dynamic ranges of luminance, which enhances their contrast in comparison to conventional low dynamic range (LDR) contents. Ultimately, the HDR imaging can render natural contrast similarly to what human perceives so as to provide realistic and immersive experience. Therefore, it is crucial to understand and measure the end users' quality of experience (QoE) of HDR contents in order to maximize QoE in various HDR applications. One such example is about how QoE of HDR contents surpasses that of conventional LDR contents. 

In general, approaches to measure QoE of multimedia contents can be divided into two categories: explicit and implicit approaches. The former refers to the traditional subjective quality assessment such as a questionnaire survey, where a number of users are asked to rate target stimuli explicitly on a given rating scale. Several subjective test methodologies have been standardized (e.g., ITU-R BT.500-13 \cite{itu500-13}) so that reliable and reproducible test results are obtained. Results of the explicit subjective quality assessment are employed as ground truth of perceptual quality in numerous studies such as development of objective perceptual quality predictors, perceptual optimization of visual signal processing systems, etc. 

In contrast, the implicit approach for measuring multimedia QoE does not require explicit rating, but observes a user's physiological response such as brain wave, blood pressure, respiration, and temperature during consumption of multimedia contents, assuming that the physiological signals convey information of the user's QoE for the given stimuli. It is expected to provide additional and complementary information in understanding human perception of contents. In addition, it has potential to enable real-time monitoring of QoE without explicit rating activities. A few recent studies have shown feasibility of such an implicit approach for compressed images \cite{Lindemann:imQ}, degraded videos \cite{Arndt:vQ}, \cite{Scholler:vQ}, \cite{Mustafa:vQ}, 3D videos \cite{Kroupi:3D-EUSIPCO}, \cite{Kroupi:3D-VPQM}, \cite{Kroupi:3D-ICME}, asynchronous audiovisual stimuli \cite{Lassalle:audiovisual}, etc. 

In this paper, we present our study of implicit QoE measurement using the electroencephalography (EEG) and peripheral physiological signals to investigate the perceptual experience of (tone-mapped) HDR videos, which has not been attempted previously to the authors' best knowledge except for our preliminary work on EEG-based measurement \cite{Moon14}. First, explicit subjective assessment is conducted to understand factors affecting QoE of tone-mapped HDR videos and obtain the ground truth for implicit quality assessment. We design the subjective assessment by considering the factors that are expected to have impact on the perceptual experience of tone-mapped HDR videos. Second, physiological signals including EEG and peripheral physiological signals are recorded during watching LDR and tone-mapped HDR videos, which are used for constructing classification systems discriminating tone-mapped HDR versus LDR and high quality versus low quality. The features significantly employed by the classification systems are also analyzed. 

On the whole, we aim at obtaining thorough understanding of perceptual experience of HDR videos as immersive multimedia. Our main contributions can be summarized as follows:

\begin{itemize}
\item The perceptual experience of tone-mapped HDR and LDR videos is implicitly analyzed based on EEG and peripheral physiological signals. These two modalities are used independently for constructing classification systems and, furthermore, fused to investigate their complementarity, from which we examine existence of difference in physiological responses with respect to subjective experience of HDR and LDR videos. Moreover, through analysis of the features employed by the classification systems, the perception of tone-mapped HDR videos are investigated and discussed in relation to physiological responses.
\item We investigate the influence of factors related to QoE of tone-mapped HDR and LDR videos, such as preference of contents and recognition of contrast. While users' preference depending on the peak luminance was evaluated in \cite{Hanhart:HDRQoE}, we consider more factors for better understanding of the elements influencing perception of HDR videos. Particularly, two kinds of subjective assessment methods, single stimulus and paired comparison methodologies \cite{itu500-13}, are employed for effective analysis of user ratings in various aspects of perceptual experience. 
\item We make public the database including recorded physiological signals, subjective assessment results, and video sequences collected in this study\footnote{http://jongseoklee.org/downloads}. In the fields of multimedia QoE and physiological signal processing, a publicly available database has a substantial role in enabling performance comparison of different studies and promoting further related studies \cite{Winkler:publicdb}, \cite{Koelstra:emotion}. However, only a few HDR video databases are currently available \cite{Eilertsen:evalvTMO}, \cite{Froehlich:HDRvdata} and HDR-associated subjective ratings and physiological signals are even rarer. We believe that our database will be valuable for further researches of relevant fields.
\end{itemize}

The rest of this paper is organized as follows. In Section \ref{sec:rw}, the concepts of the HDR imaging and tone-mapping are briefly described, and preceding studies on implicit measurement of human perception are reviewed. The experimental materials and procedure are explained in Section \ref{sec:exp}. Section \ref{sec:sub} describes the results of the subjective rating analysis. The pre-processing of physiological signals and classification schemes are described in Section \ref{sec:cls}, and classification results follow in Section \ref{sec:res}. Finally, the conclusion is given in Section \ref{sec:concl}.

\section{Related work}
\label{sec:rw}
\subsection{HDR and tone-mapping}
\label{subsec:rw.HDR}

From starlight to sunlight, the nature contains a very wide range of luminance. But conventional LDR digital images typically have 8 bit-depth, which means that they represent luminance on 256 steps, from 0 to 255. Thus, pixel values that exceed this range are replaced by 0 or 255, which gives rise to detail loss. The HDR imaging reduces this contrast loss by allocating more bits to express luminance, in other words, extending the dynamic range of luminance. Therefore, it is expected to better preserve contrast in natural scenes that the human visual system (HVS) captures. 

The color perception of LDR and HDR contents was investigated in \cite{Kim:colorperception}. Subjects were required to estimate the lightness, hue, and colorfulness of color patches for different peak luminance levels. The increase of the perceived lightness mainly appeared on medium light colors, which indicates that the shape of the perceived lightness curve changes in high peak luminance. Also, the colorfulness mainly increased in bright light colors.

Preference of HDR videos was evaluated in \cite{Hanhart:HDRQoE}. Eight original videos were tone-mapped with peak luminance values of 100 $cd/m^2$, 400 $cd/m^2$, 1,000 $cd/m^2$, and 4,000 $cd/m^2$, where the first and last cases correspond to conventional LDR and HDR videos, respectively. It was shown that the QoE increases as the peak luminance value increases, thus the QoE is maximized at 4,000 $cd/m^2$.

Since HDR images use more than 8 bits for each color component, conventional LDR display devices cannot properly display them, and HDR displays are not available yet in common. Therefore, the pixel values of a HDR image need to be adjusted so that they lie within the dynamic range of common LDR devices for display, which is called tone-mapping. There exist many studies to develop tone-mapping operators (TMOs) in literature \cite{Devlin:tmosurvey}, which can be divided into two broad categories, namely, global operators and local operators. Global operators use fixed mapping functions for all pixels in an image. Therefore, they are independent of local values as opposed to local operators that depend on features extracted from surrounding pixels. Global operators are usually simpler and faster than local operators, but local operators can preserve better local contrast to which HVS is sensitive. Also, in regard to the type of target contents, TMOs can be categorized into image TMOs and video TMOs. Operations of image TMOs and video TMOs are not different essentially, but the artifacts on the temporal axis such as flickering, ghosting, and temporal inconsistency of brightness have to be considered in video TMOs.

In \cite{Kuang:TMOeval}, the rendering accuracies of six state-of-the-art TMOs have been evaluated in comparison to the corresponding real-world scenes. The results showed that the TMOs facilitate reproduction of real-world scenes in LDR displays with high accuracy, especially in terms of the overall contrast, contrast in shadow, and colorfulness in shadow. 

Considering the practical environments and notable performance of TMOs, tone-mapped HDR videos are used in this work to evaluate perceptual quality of HDR contents. In particular, the video TMO by Boitard et al. \cite{Boitard:vTMO} is used for tone-mapping of our HDR videos, where the TMO by Reinhard et al. \cite{Reinhard02:TMO}, which is popularly used in several studies (e.g., \cite{Kang:vTMO}, \cite{Ramsey:vTMO}), is employed as a base image TMO.


\subsection{Implicit QoE measurement}
\label{subsec:rw.imp}

\begin{table*}[!t]
\renewcommand{\arraystretch}{1.3}
\centering
\caption{Summary of implicit QoE measurement studies}
\label{table:prestudies}
\begin{tabular}{>{\centering}m{0.7cm}|>{\centering}m{2cm}|>{\centering}m{1.4cm}|>{\centering}m{1.5cm}|>{\centering}m{1cm}|>{\centering}m{4cm}|m{4cm}} \hline 
\rule{0in}{2ex}
 & Channel & Type of media & Quality factor & Number of subjects & Scheme & \centering{Results} \tabularnewline \hline \hline
\cite{Antons:sQ} & EEG & Speech & Presence of artifacts & 10 & Subject-dependent classification for existence of quality changes & Balanced accuracy over 90\% \tabularnewline \hline
\cite{Lindemann:imQ} & EEG& Image & Level of JPEG artifacts & 10 & Correlation between appearance of artifacts and EEG & Changes of EEG in the occipital area under existence of JPEG artifacts \tabularnewline \hline
\cite{Arndt:vQ} & EEG & Video & Level of blockiness artifacts & 10 & ANOVA for effect of blockiness on the P300 amplitude & Significant difference ($p < 0.02$) \tabularnewline \hline
\cite{Scholler:vQ} & EEG & Video & Quality changes & 9 & Correlation between quality changes and the P300 amplitude & Correlation coefficient of 0.89 \tabularnewline \hline
\cite{Mustafa:vQ} & EEG & Video & Presence, severity, and type of artifacts & 8 & Classification of artifact presence, severity, and type & Classification accuracy of 85\% (presence), 94\% (severity), and 64\% (type) \tabularnewline \hline
\cite{Kim:3Dvfatigue-EEG} & EEG & 3D Video & Visual fatigue & 5 & T-test for difference of the beta band power between 2D and 3D videos & Significant difference ($p$ = 0.028) \tabularnewline \hline
\cite{Li:3Dvfatigue} & EEG & 3D Video & Visual fatigue & 7 & Correlation between the watching duration and EEG & Increase of the beta band power and P700 delay according to increase of the watching duration \tabularnewline \hline
\cite{Kroupi:3D-EUSIPCO} & EEG & 3D Video & Subjective QoE & 16 & Correlation between QoE and EEG & Activation of the alpha band in the right frontal lobe for low QoE; Activation of the beta band in the right parietal lobe for high 3D QoE \tabularnewline \hline
\cite{Kim:3Dvfatigue-fMRI} & fMRI & 3D Video & Visual fatigue & 5 & T-test for difference of the brain activity for small and large visual depths & High t-value in the frontal eye field \tabularnewline \noalign{\hrule height 1pt}
\cite{Kroupi:3D-VPQM} & EEG, heart rate, respiration & 3D Video & 2D vs. 3D & 16 & Subject-independent classification between 2D and 3D videos & Classification accuracy of 54.69\% (EEG) \tabularnewline \hline
\cite{Kroupi:3D-ICME} & EEG, heart rate, respiration & 3D Video & Level of sensation of reality & 16 & Subject-dependent classification for low vs. high levels of sensation of reality & Matthews correlation coefficients of 0.65 (EEG) and 0.16 (heart rate, respiration) \tabularnewline \hline
\cite{Lassalle:audiovisual} & Blood volume pulse, temperature, skin conductance, eye tracking & Audiovisual stimuli & Degradation of quality & 34 & ANOVA for quality effect on physiological signals & Significant difference ($p < 0.01$) for skin conductance \tabularnewline \hline
\end{tabular}
\end{table*}

As aforementioned, the implicit approach to measure QoE does not require explicit rating, but monitors physiological signals such as brain wave, blood pressure, respiration, temperature, and so on. There are a couple of merits of the implicit approach; it is easy to obtain temporally continuous data for real-time monitoring and has potential to decrease any subjective bias in rating. 
Table \ref{table:prestudies} lists recent studies of implicit QoE measurement. It can be seen that different channels for implicit measurement have been explored for QoE assessment. 

The brain activity is the most informative physiological signal for monitoring perceptual experience of multimedia. It is typically measured by EEG, and sometimes functional magnetic resonance imaging (fMRI) is also used \cite{Kim:3Dvfatigue-fMRI}. The EEG system is comparatively cheap, space-saving, and wearable, and has a high temporal resolution (milliseconds). For these reasons, EEG has been mainly employed by numerous studies.

In \cite{Antons:sQ}, it was examined whether the existence of quality changes in speech signals can be detected based on EEG. Results showed that the P300 response, which is a positive peak signal of EEG occurring approximately 300 ms after a stimulus, is delayed for small degradations, and stronger degradations cause higher amplitudes of the P300 component.

Much more studies have been conducted for visual stimuli than acoustic stimuli. In \cite{Lindemann:imQ}, the quality of JPEG compressed images was evaluated using EEG. It was observed that the EEG signals recorded from the occipital area (O1, Oz, and O2 electrodes in the international 10-20 system) change when JPEG artifacts appear.

EEG responses with respect to video quality were monitored in \cite{Arndt:vQ}, \cite{Scholler:vQ}, \cite{Mustafa:vQ}.
The amplitude of the P300 component showed high correlation with the differential mean opinion score (DMOS) in \cite{Arndt:vQ} and the change detection rate of subjects in \cite{Scholler:vQ}.
In \cite{Mustafa:vQ}, it was shown that EEG can be used to classify presence, strength, and types of artifacts in videos, such as popping, popping on persons, blurring, blurring on persons, and ghosting on persons.

There exist studies investigating detection of visual fatigue due to 3D stimuli through EEG.
In \cite{Kim:3Dvfatigue-EEG}, it was shown that the power of the EEG beta band increases during watching 3D videos causing stronger visual fatigue than 2D videos.
A similar observation was obtained in \cite{Li:3Dvfatigue}, i.e., according to the increase of the watching duration of visual stimuli, the power of the beta band increases. In addition, it was observed the P700 response is delayed according to the increase of the watching duration. These tendencies were shown in both cases of 2D and 3D stimuli, but stronger in case of 3D stimuli.

QoE of 3D videos was explored in \cite{Kroupi:3D-EUSIPCO} using EEG. It was observed that the EEG alpha band in the right frontal lobe is activated when perceived quality is low. In addition, the right parietal lobe, which is relevant to positive emotional processes, is activated when QoE of 3D videos is high. 

There have been attempts to explore the relationship between multimedia quality perception and peripheral physiological signals.
In \cite{Kroupi:3D-VPQM} and \cite{Kroupi:3D-ICME} the perception of 3D videos was measured by heart rate and respiration as well as EEG. Both EEG and peripheral physiological signals were useful for subject-dependent classification of high vs. low sensation of reality \cite{Kroupi:3D-ICME}. However, subject-independent classification between 2D and 3D videos was not feasible based on peripheral physiological signals, unlike EEG \cite{Kroupi:3D-VPQM}.

In \cite{Lassalle:audiovisual}, blood volume pulse, temperature, skin conductance, and eye tracking were employed to explore QoE of audiovisual stimuli. Thirty four subjects evaluated the video quality, audio quality, and overall quality of three audiovisual contents with three different types of quality degradation: no quality loss, bitrate reduction, and asynchronous artifacts. It was possible to identify the degradation of audiovisual stimuli only for the skin conductance ($p < 0.01$). However, this result remains inconclusive because of limitation of the experimental protocol.

From these studies, the potential of EEG and peripheral physiological signals for evaluating QoE is verified for various media. Generally, EEG performs better in distinguishing perceptual experience than peripheral physiological signals as shown in \cite{Kroupi:3D-VPQM}, \cite{Kroupi:3D-ICME}. Moreover, although a few studies showed significant performance under subject-independent scenarios, it is still difficult to overcome individual differences in physiological signal analysis.

\section{Experiment}
\label{sec:exp}

\subsection{Video stimuli}
\label{subsec:exp.v}

\newcommand{\rb}[1]{\raisebox{-11ex}[0pt]{#1}}
\begin{table*}[!t]
\renewcommand{\arraystretch}{1.3}
\caption{Video stimuli used in the experiment}
\label{table:v}
\centering
\begin{tabular}{>{\centering}m{1cm}|>{\centering}m{3cm}|>{\centering}m{3cm}|>{\centering}m{4.5cm}|>{\centering}m{1cm}}
\hline
\multirow{2}*[-.3ex]{Content} & \multicolumn{2}{c|}{Thumbnail} & \multirow{2}*[-.3ex]{Description} & \multirow{2}*[-.3ex]{Length} \tabularnewline \cline{2-3} 
& LDR & tone-mapped HDR & & \tabularnewline \hline \hline 
\centering hall &
\rb{
	\includegraphics[width=3cm,height=1.69cm]{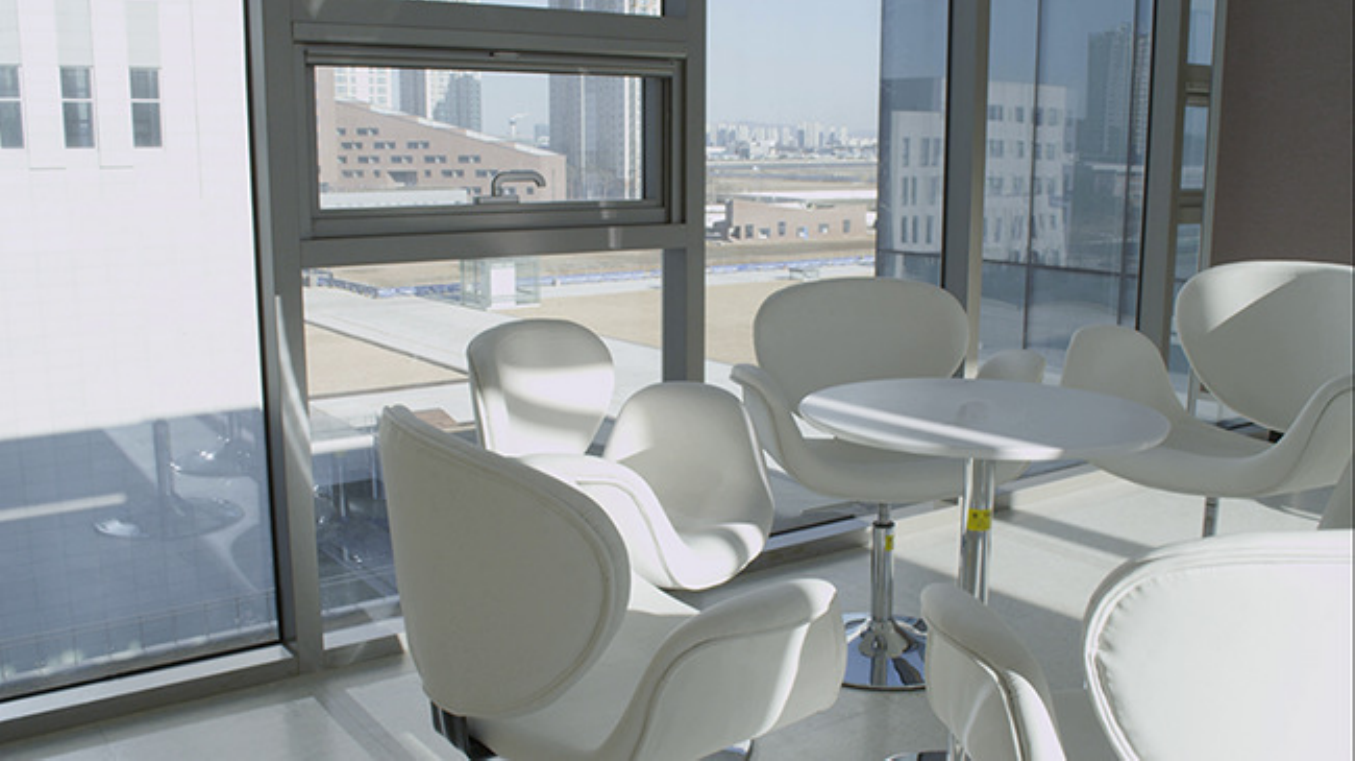}} & 
\rb{
	\includegraphics[width=3cm,height=1.69cm]{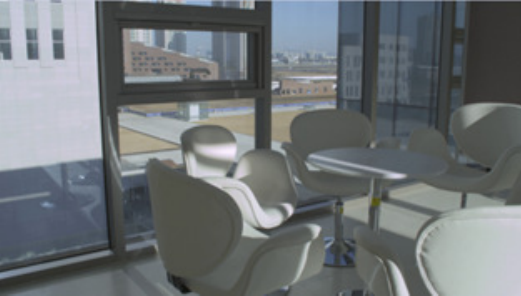}} & Hall with a glass wall and metallic objects including chairs and a table. A horizontal pan shot. & \raisebox{-5ex}[0pt]{40 sec} \tabularnewline[11ex] 
\hline
objects & 
\rb{
	\includegraphics[width=3cm,height=1.69cm]{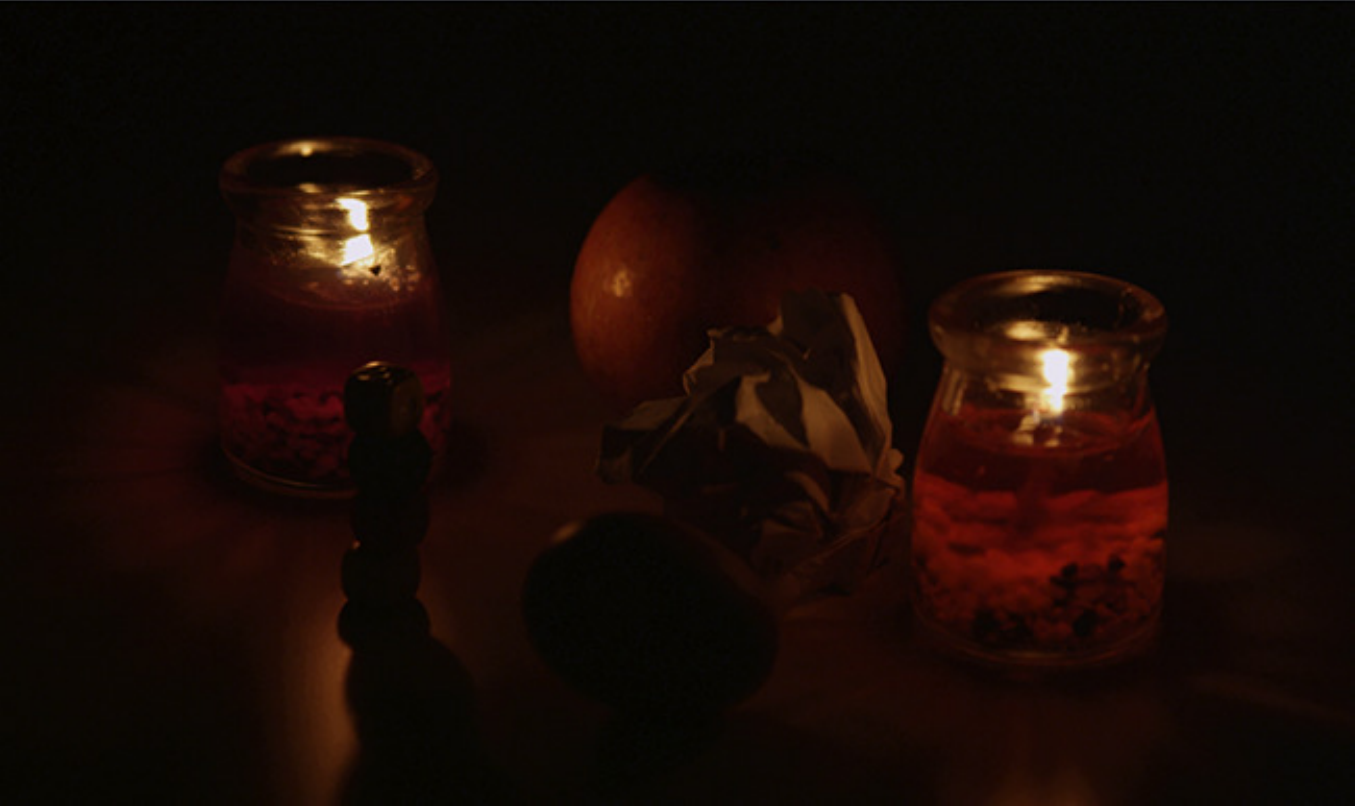}} & 
\rb{
	\includegraphics[width=3cm,height=1.69cm]{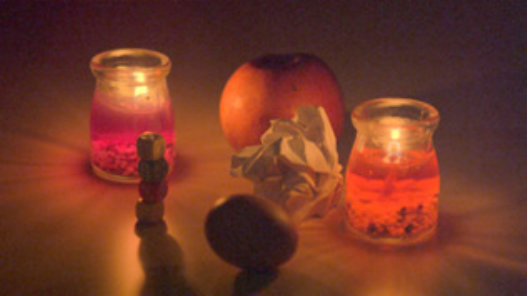}} & Small objects in a dark room with candles. & \raisebox{-5ex}[0pt]{50 sec} \tabularnewline[11ex] 
\hline
sky & 
\rb{
	\includegraphics[width=3cm,height=1.69cm]{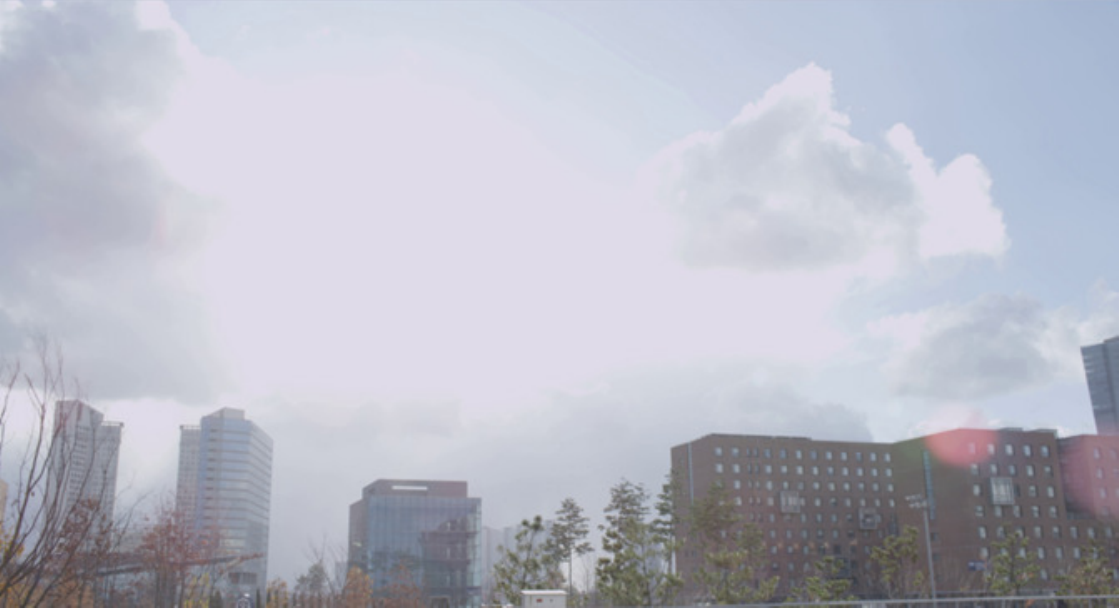}} & 
\rb{
	\includegraphics[width=3cm,height=1.69cm]{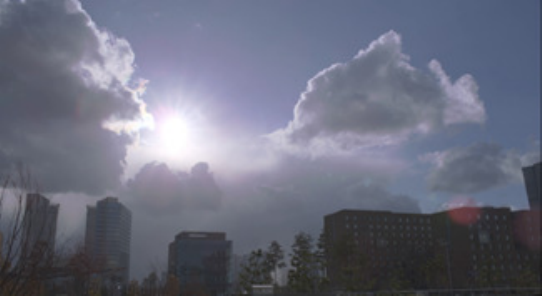}} & Clouds drifting in the sky. & \raisebox{-5ex}[0pt]{30 sec} \tabularnewline[11ex] 
\hline
window & 
\rb{
	\includegraphics[width=3cm,height=1.69cm]{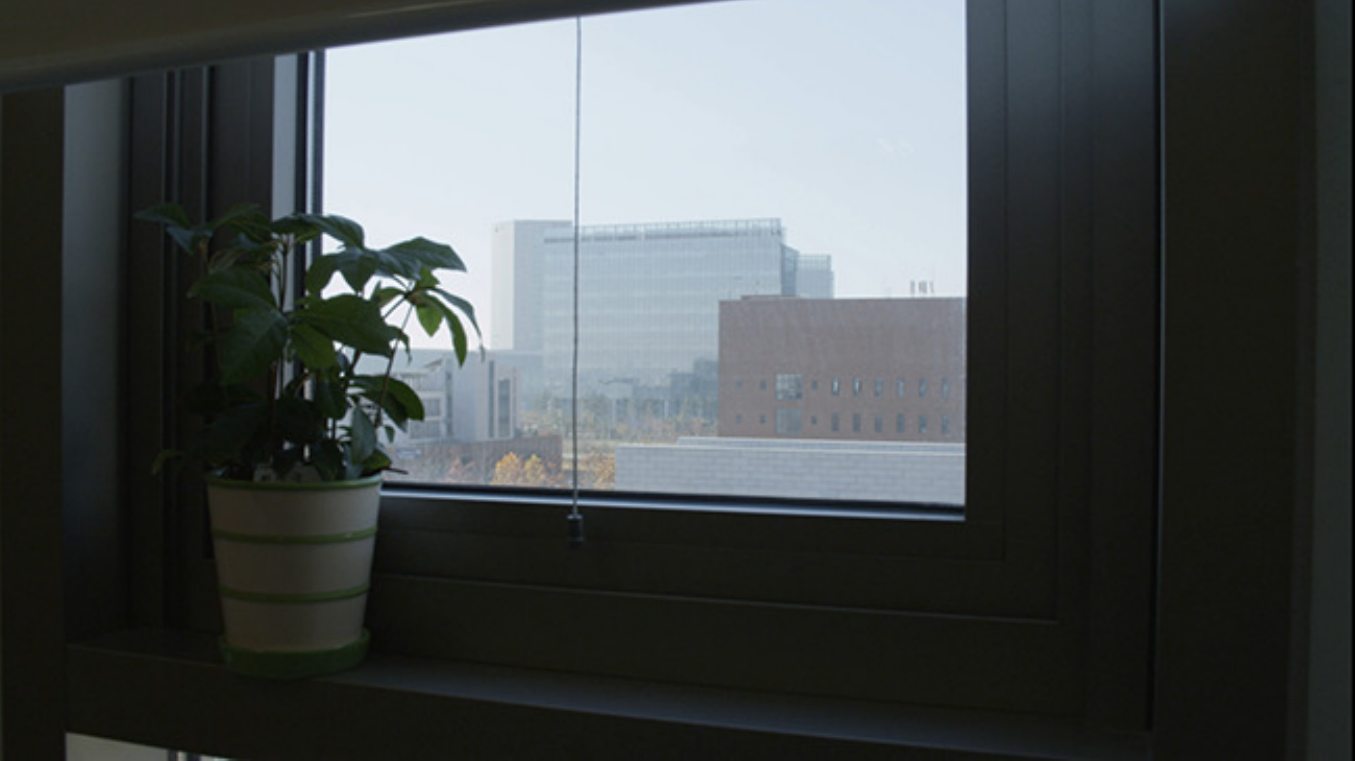}} & 
\rb{
	\includegraphics[width=3cm,height=1.69cm]{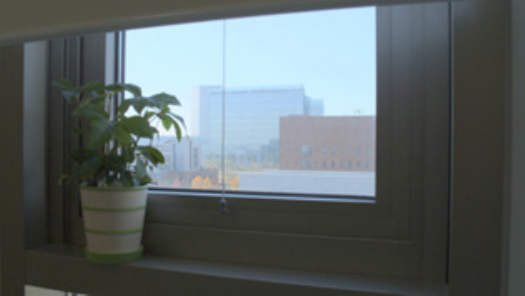}} & Window with a pot. Brightness changes by a window blind. & \raisebox{-5ex}[0pt]{20 sec} \tabularnewline[11ex]
\hline
\end{tabular}
\end{table*}

Five video sequences were captured by a RED EPIC camera using its HDRx mode. One is used for training subjects, and the other four for test. The sequences contain various indoor and outdoor scenes. The details of the contents are described in Table \ref{table:v} with thumbnails. Using the REDCINE-X PRO software, HDR and LDR sequences having a full HD resolution (1920$\times$1080 pixels) and a frame rate of 30 fps were extracted from the raw videos. Then, the HDR sequences were tone-mapped using the video TMO by Boitard et al. \cite{Boitard:vTMO} with the image TMO by Reinhard et al. \cite{Reinhard02:TMO} for frame-wise processing. The tone-mapping process was conducted using the Banterle's HDR Toolbox \cite{Banterle:HDRtb}.

\subsection{Subjects}
\label{subsec:exp.sub}

Five subjects, three of which are males, participated in the experiment. They were between 23 and 34 years old. All subjects had normal or corrected-to-normal vision. They had not experienced HDR or tone-mapped HDR contents.

\subsection{Procedure}
\label{subsec:exp.proc}

First, physiological signal sensors were equipped to the subjects. Then, the subjects sat comfortably at a distance of 2.5 times the height of monitor from the screen, but they were instructed to move as little as possible to minimize movement artifacts in the recorded EEG. The training session was conducted using the training video sequence to explain the procedure of the experiment.

The test video sequences were displayed on a 84-inch LCD monitor having a native resolution of 1920$\times$1080 pixels. A gray screen was shown for ten seconds before each video sequence was shown. The physiological signals recorded in this duration are considered as the baseline signals. The tone-mapped HDR and LDR video sequences were shown on the screen in a randomized order. To avoid direct comparison between tone-mapped HDR and LDR videos, it was made sure that the same contents were not displayed consecutively. The physiological signal recording took six minutes for each subject.

After the recording of physiological signals, the video sequences were shown again as tone-mapped HDR and LDR pairs of the same contents for subjective rating. After each pair was shown, the subjects were asked to rate the two videos. The order of the contents among four and the order between the tone-mapped HDR and LDR videos in each pair were all randomized. The subjective rating took roughly 6 minutes with slight variations in the time for rating across subjects.

\subsection{Physiological signal measurement}
\label{subsec:exp.phy}

\begin{figure}[!t]
	\centering
	\includegraphics[width=2.5in]{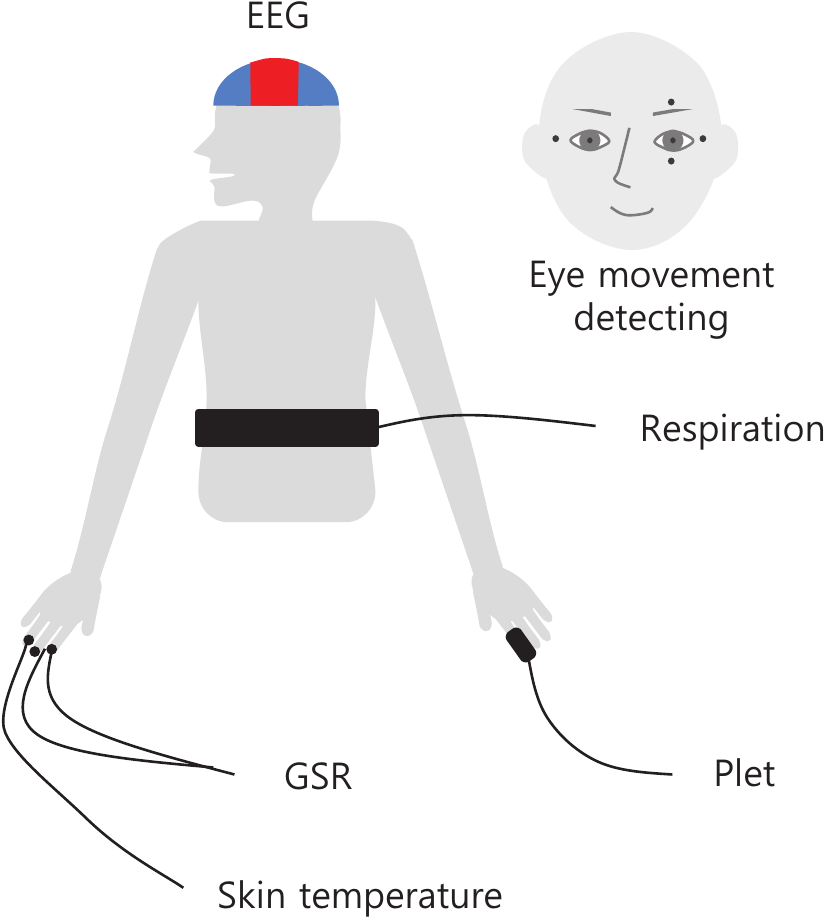}
	\caption{Placement of physiological signal sensors}
	\label{fig:equip}
\end{figure}

The EEG, galvanic skin response (GSR), plethysmography, respiration, and skin temperature were recorded as physiological cues about perceptual experience of subjects. The signals were recorded by a Biosemi ActiveTwo system. Figure \ref{fig:equip} illustrates the placement of the physiological signal sensors.

\subsubsection{EEG}
The EEG signal was recorded by 32-channel electrodes. Their positions on the scalp were fixed by electrode holders on a electrode cap. The subjects wore a medium or large size electrode cap, and electrodes were connected to the surface of the scalp through highly conductive gel. Four additional electrodes were placed around eyes for rejecting eye blinking and movement artifacts. 

\subsubsection{GSR}
The GSR, also known as skin conductance, indicates the state of the sympathetic nervous system that controls the amount of sweat on the skin. When the sympathetic nervous system arouses, the amount of sweat increases, which in turn causes decrease of the electric resistance of skin. Two electrodes were attached to the index finger and middle finger of the right hand.

\subsubsection{Plethysmography}
The plethysmography, a method to measure changes in volume, was applied to evaluate the heart rate relevant physiological features. A plethysmograph sensor (MLT1020 of ADI Instruments) was mounted on the index finger of the left hand. 

\subsubsection{Respiration}
A Nihon Kohden TR-753T respiration belt made of latex was employed to monitor respiration. The belt was tied on the abdomen without interrupting natural abdominal movement and causing difficulty in breathing. 

\subsubsection{Skin temperature}
A high precision temperature sensor (Agilent 21078A) was used to measure skin temperature. The sensor was attached to the ring finger of the right hand.

\subsection{Questionnaire survey}
\label{subsec:exp.Q}

The subjective rating form consisted of two types of questions. One includes three single-stimulus questions for each video \cite{itu500-13}, which ask quality of contrast (Q1), interest to contents (Q2), and overall quality (Q3). A 9-point scoring system visualized by self-assessment manikins (SAM) \cite{Bradley:SAM} was employed for these questions. 
The rating scales range from  bad quality (1) to good quality (9) for Q1, uninteresting (1) to interesting (9) for Q2, and bad quality (1) to good quality (9) for Q3.

The other comprises a paired comparison type of questions for direct pair-wise comparison between the tone-mapped HDR and LDR videos of the same contents \cite{itu500-13}. Two questions were used, one for comparing quality (compQ1) and the other for comparing naturalness (compQ2). A 7-scale scoring system was used for these questions, which spans ``the first video is significantly better'', ``the first video is better'', ``the first video is slightly better'', ``same'', ``the second video is slightly better'', ``the second video is better'', and ``the second video is significantly better''. Each rating is converted into a numeric value for analysis in such a way that a positive (or negative) value is assigned if tone-mapped HDR (or LDR) video is preferred ($\pm$1 for ``slightly better'' and $\pm$3 for ``significantly better'').

\section{Subjective rating analysis}
\label{sec:sub}

\begin{table}[!t]
\renewcommand{\arraystretch}{1.3}
\centering
\caption{Summary of subjective ratings for single-stimulus tests}
\label{table:ss}
\begin{tabular}{c|c c|c c|c} \hline
\rule{0in}{2ex}
\multirow{2}*[-.3ex]{Question}&\multicolumn{2}{c|}{LDR}&\multicolumn{2}{c|}{Tone-mapped HDR}&\multirow{2}*[-.3ex]{$p$-val}\\ \cline{2-5}
\rule{0in}{2ex}
& mean & std & mean & std &\\ \hline \hline
\rule{0in}{2ex}
Q1 (contrast) &5.85&2.01&6.65&2.03&0.2181\\
Q2 (content) &5.80&2.07&7.00&1.21&0.0326\\
Q3 (quality) &5.70&1.92&6.90&1.62&0.0394\\
\hline
\end{tabular}
\end{table}

\begin{table}[!t]
\renewcommand{\arraystretch}{1.3}
\centering
\caption{Pearson correlation coefficients between subject rating questions}
\label{table:PCC}
\begin{tabular}{>{\centering}m{2cm}|>{\centering}m{2cm} | >{\centering}m{2cm}} \hline
\rule{0in}{2ex}
 & Q1 (contrast) & Q2 (content) \tabularnewline \hline \hline \rule{0in}{2ex}
Q2 (content) & 0.5450 & - \tabularnewline \hline \rule{0in}{2ex}
Q3 (quality) & 0.8145 & 0.8395 \tabularnewline
\hline 
\end{tabular}
\end{table}

\begin{table}[!t]
\renewcommand{\arraystretch}{1.3}
\centering
\caption{Result of subjective ratings for paired comparison tests}
\label{table:pc}
\begin{tabular}{>{\centering}m{3.2cm}|>{\centering}m{3.2cm}|c} \hline
\multicolumn{2}{c|}{Mean}&\multirow{2}*[-.3ex]{PCC}\\ \cline{1-2}
\rule{0in}{2ex}
compQ1 (quality) &compQ2 (naturalness)& \\ \hline \hline
\rule{0in}{2ex}
0.45&0.10&0.7223\\
\hline
\end{tabular}
\end{table}

In this section, the results of subjective rating are described with the aim of understanding characteristics of QoE of tone-mapped HDR videos and analyzing factors affecting QoE of tone-mapped HDR videos. 

The result of the single-stimulus subjective tests is shown in Table \ref{table:ss}. The table presents the mean and standard deviation values of the scores, and the $p$-values of the t-tests between the scores of the tone-mapped HDR and LDR videos are also shown, where the number of samples for a t-test is 20 ($=$5 subjects$\times$4 contents) in each group. For Q1, the tone-mapped HDR videos obtained an average score of 6.65, which is higher than the score of the LDR videos (5.85), but the difference is not statistically significant ($p$ $>$ 0.05).
For Q2, the tone-mapped HDR videos obtained a score of 7.00 and, surprisingly, this is significantly higher ($p$ $<$ 0.05) than that of the LDR videos (5.80). This indicates that HDR imaging can make videos more interesting, even if the content does not change. For Q3, the tone-mapped HDR videos acquired a score of 6.90 on average, which is significantly higher ($p$ $<$ 0.05) than that of the LDR videos (5.70), as expected. 
 
The insignificance of the score difference between tone-mapped HDR and LDR for Q1 may be considered contradictory, because the significantly better quality of tone-mapped HDR (shown in Q3) is presumably due to the enhanced contrast. Our guess is that, although the subjects perceived difference between the tone-mapped HDR and LDR videos, some subjects did not realize that the difference is originated from the enhanced contrast in tone-mapped HDR because it is rather a technical concept.

To examine the relationship between the scores for Q1, Q2, and Q3, the Pearson correlation coefficient (PCC) with 40 samples ($=$5 subjects$\times$8 videos) is computed for each pair of questions and shown in Table \ref{table:PCC}. Q3 shows high correlation with both Q1 and Q2, but the correlation between Q1 and Q2 is not very high.

The mean score values of the paired comparison questions are shown in Table \ref{table:pc}. For both questions, the mean values are positive, indicating the tone-mapped HDR videos have better quality and are more natural. A relatively lower score is obtained for compQ2 than compQ1. This is because the tone-mapped HDR videos are perceived as artificial in some cases, as subjects  are familiar to conventional LDR videos. The PCC between compQ1 and compQ2 is 0.7223, which indicates that the naturalness is also attached to the quality.

\section{EEG and peripheral signals analysis}
\label{sec:cls}

In this section, the classification between tone-mapped HDR and LDR videos is implemented to explore whether the recorded physiological signals can be used for assessing the perception of tone-mapped HDR videos. Also, the relationship between the physiological signals and subjective rating is investigated for further understanding of the QoE of tone-mapped HDR videos.

\begin{table*}[!t]
\renewcommand{\arraystretch}{1.3}
\centering
\caption{Peripheral signal features extracted for classification}
\label{table:feature}
\begin{tabular}{>{\centering}m{4.5cm}|>{\centering}m{2.5cm}|l} \hline
Sensor & Feature & \multicolumn{1}{c}{Description} \\ \hline \hline
\multirow{4}*[-.3ex]{Galvanic skin response (GSR)} & GSR\_M & Mean of GSR signals \\ \cline{2-3}
& GSR\_Std & Standard deviation of GSR signals \\ \cline{2-3}
& GSR\_derM & Mean of differentiated GSR signals \\ \cline{2-3}
& GSR\_derStd & Standard deviation of differentiated GSR signals \\ \hline
\multirow{3}*[-.3ex]{Respiration} & Resp\_derM & Mean of differentiated respiration signals \\ \cline{2-3}
& Resp\_Std & Standard deviation of respiration signals \\ \cline{2-3}
& Resp\_peaktM & Mean of peak to peak time \\ \hline
\multirow{4}*[-.3ex]{Plethysmography (Plet)} & Plet\_HRM & Mean of heart rate \\ \cline{2-3}
& Plet\_HRStd & Standard deviation of heart rate \\ \cline{2-3}
& Plet\_HRVM & Mean of temporal variation of heart rate \\ \cline{2-3}
& Plet\_HRVStd & Standard deviation of temporal variation of heart rate \\ \hline
\multirow{2}*[-.3ex]{Skin temperature} & Temp\_M & Mean of skin temperature signals \\ \cline{2-3}
& Temp\_derM & Mean of differentiated skin temperature signals \\
\hline
\end{tabular}
\end{table*}

\subsection{Signal pre-processing}
\label{subsec:cls.sp}

First, all physiological signals are downsampled to 256 Hz and one-second signals are removed both at the beginning and the end of each period to obtain stabilized signals only.

The EEG signals are bandpass-filtered within 2-100 Hz and the vertex electrode Cz is employed as a reference channel. The artifact signals such as eye blinking and eye movement are rejected with independent component analysis (ICA) \cite{Bell:ICA}, \cite{Lee:ICAextd}. The respiration signals are also bandpass-filtered between 0-10 Hz to remove noise. The pre-processing is conducted by using the EEGLAB ToolBox \cite{Delorme:EEGLAB}.

In the previous work \cite{Chanel:emotion}, \cite{Khalili:emotion}, \cite{Konopka:mvEEG}, the physiological signals divided into 6 to 12.5 seconds have been used for analysis. It arises from the consideration about rapid changes of the signal characteristics, especially for EEG. While the temporal change of other peripheral signals is not as rapid as EEG, it was shown that perceptual states are reasonably well recognized by the peripheral signals in a few seconds period \cite{Chanel:emotion}, \cite{Khalili:emotion}. Therefore, in this work, the recorded physiological signals are divided into 10 second-long segments, and thus total 28 segments are obtained for each subject.

\subsection{Feature extraction}
\label{subsec:cls.f}

To extract features, the frequency power of the EEG is calculated by the Welch's method with a 256 samples window. Then, the baseline power is subtracted from the stimulus power, and the powers in theta (3-7 Hz), alpha (8-13 Hz), beta (14-29 Hz) and gamma (30-47 Hz) frequency bands are extracted for each electrode channel. Therefore, total 124 features (4 frequency bands $\times$ 31 electrodes, except for the reference channel Cz in the 32-channel system) are obtained from the EEG signals. 

The features extracted from the recorded peripheral physiological signals are summarized in Table \ref{table:feature}. From the GSR, its mean and standard deviation values are obtained, and the mean and standard deviation of the differentiated GSR are also calculated. Regarding the respiration signals, the average peak-to-peak time, standard deviation and mean of the differentiated respiration signals are acquired. The heart rate and its temporal variation are obtained from the plethysmography signals using the real-time QRS detection algorithm \cite{Pan:QRS}, and their mean and standard deviation values are extracted as features. Finally, the mean values of the skin temperature signals and their derivatives are extracted.

\subsection{Classification scheme}
\label{subsec:cls.cls}

We consider two different binary classification problems. The first is to classify tone-mapped HDR and LDR, where classification performance is evaluated in terms of classification accuracy. In the subject-dependent scenario, for each subject, a classification system is trained using 27 segments among the 28 segments obtained as a result of segmentation, and the remaining one segment is used for classification test, called a trial, which is repeated until all segments are tested. In the subject-independent scenario, the signals of one subject is used for testing among the five subjects, and the remaining four subjects' signals are used for training a classification system, which is considered as a trial; this is repeated five times until all subjects are tested.

The second is classification between high quality and low quality. Both the contrast quality (Q1) and the overall quality (Q3) are considered. The stimuli rated higher than 5, which means definitely high quality, form one class and the other stimuli form the other class. Subject 1 was excluded in this classification problem, because of its extreme data imbalance between the two classes.
Both subject-dependent and subject-independent scenarios are considered as in the first classification problem. The classification performance is evaluated in terms of F-measure and balanced accuracy because of the imbalance of the class distribution. The F-measure normally considers the positive class more importantly than the negative class, whereas both classes have to be considered equally in our case. Therefore, the F-measure is calculated two times with changing the positive and negative classes and their average is considered as the F-measure of the whole classification. The balanced accuracy \cite{Velez:balacc} is defined as the average of the recalls of both classes. 
When class $A$ and class $B$ exist, the balanced accuracy ($balacc$) is defined as follows: 

\begin{equation}
\label{eq:bal}
balacc = \frac{1}{2}\cdot \left(\frac{true A}{number of A} + \frac{true B}{number of B}\right)
\end{equation}
where  $number of A$ and $number of B$ indicate the sizes of class A and class B, respectively, and $true A$ and $true B$ represent the numbers of data classified correctly for the two classes.

Since a large number of features are extracted through the feature extraction process, we conduct feature selection for efficient training of classifiers. For this, the features are ranked according to their significance on classification for the training data. The significance is evaluated by the Fisher's linear discriminant analysis, which is given by

\begin{equation}
\label{eq:Fisher}
J(f) = \frac{\| \mu_A - \mu_B \|}{\sigma^2_A + \sigma^2_B}
\end{equation}
where $\mu_i$ and $\sigma_i$ are the mean and standard deviation of feature $f$ for class $i$, respectively.
The numerator indicates the scatter between class $A$ and $B$, and the denominator implies the scatter within classes. Therefore, a larger value of $J(f)$ means feature $f$ has more significance for distinguishing class $A$ and class $B$.

We employ neural networks having one hidden layer with sigmoid neurons as classifiers. The neural networks are trained by the Levenberg-Marquardt algorithm, which is one of the fastest and best methods for training \cite{More:LM}. It is necessary to optimize the number of hidden neurons for the best performance. These parameters are optimized for each trial using a grid search through leave-one-out cross-validation in the training dataset. These classification schemes are conducted by using EEG and peripheral signals separately. Finally, the average performance of the classifiers is calculated from independent 10 repetitions with different random seeds for initializing the weights of the neural networks.

\subsection{Classifier fusion}
\label{subsec:cls.clsfusion}

To examine complementarity of EEG and peripheral physiological signals, we employ a late fusion method. Classification results obtained from the two modalities are fused by a weighted product method, whose decision rule is written as \cite{Koelstra:emotion}, \cite{Lee:audiovisual}.

\begin{equation}
\label{eq:Fisher}
c = \arg\max_{i} \left\lbrace P_i(X|\lambda_{EEG})^w \times P_i(X|\lambda_{peri})^{1-w}\right\rbrace
\end{equation}
where $P_i(X|\lambda)$ is the output (probability) of given data $X$ for the $i$th class by classifier $\lambda$, $\lambda_{EEG}$ and $\lambda_{peri}$ are the trained classifiers for EEG and peripheral signals, respectively, and $w$ is the fusion weight between 0 and 1.
The weight is optimized for each trial by using a grid search through leave-one-out cross-validation in the training dataset. 

\begin{figure}[!t]
	\centering
	\subfloat[]{
		\includegraphics[width=2.2in]{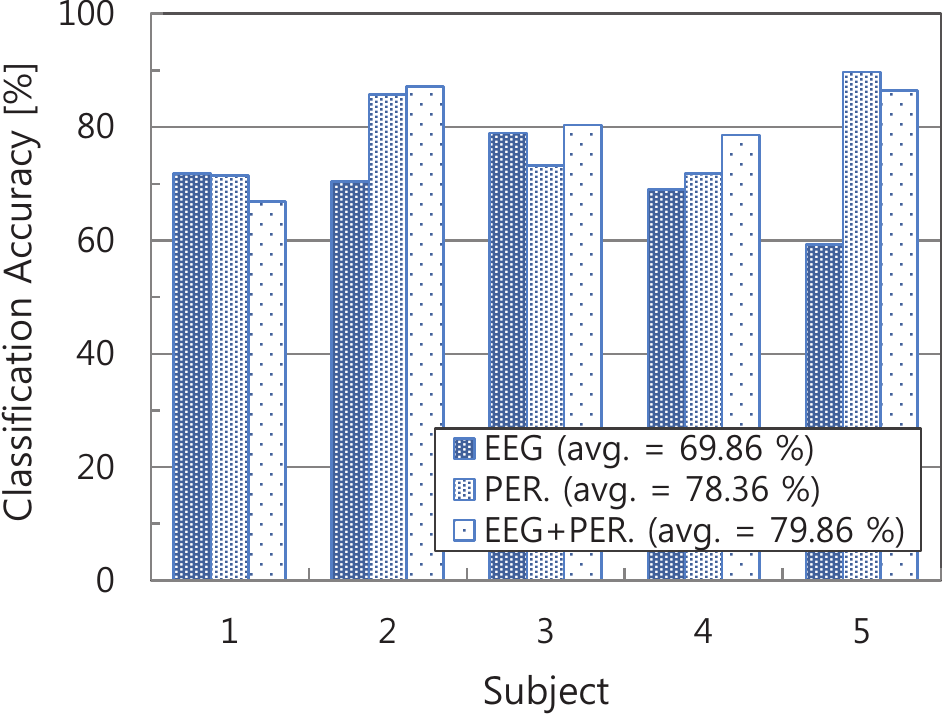}}
	
	\vspace{0.2in}
	
	\subfloat[]{
		\includegraphics[width=2.2in]{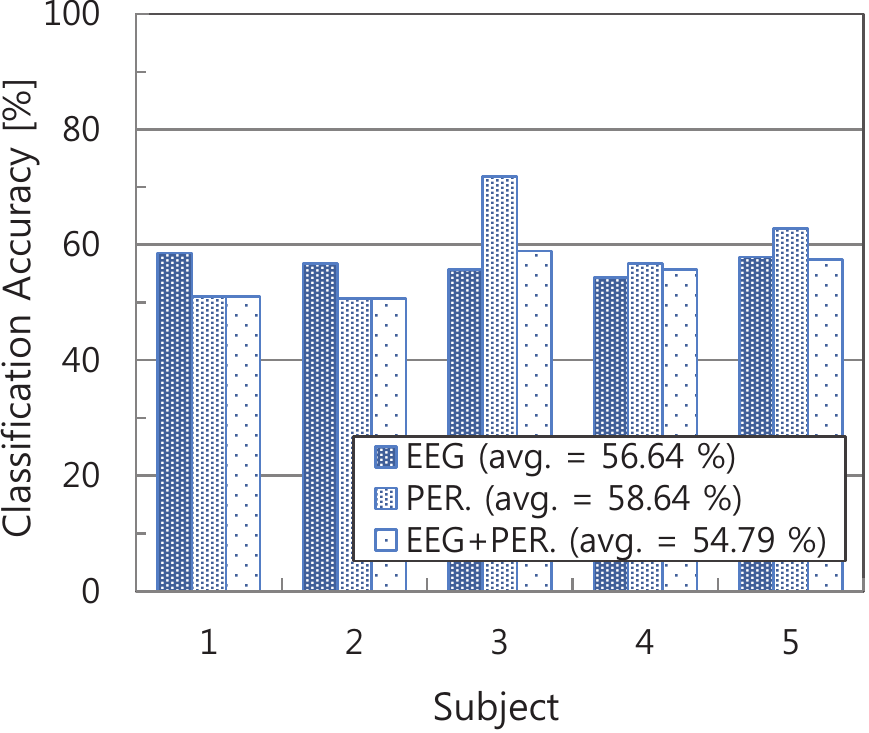}}
	\caption{Results of tone-mapped HDR vs. LDR classification for (a) the subject-dependent scenario, and (b) the subject-independent scenario}
	\label{fig:HL}
\end{figure}

\section{Results}
\label{sec:res}

\subsection{Tone-mapped HDR versus LDR classification}
\label{subsec:res.HL}

The results of the classification between tone-mapped HDR and LDR videos are shown in Figure \ref{fig:HL}. In the subject-dependent scenario, average classification accuracies of 69.86\%, 78.36\%, and 79.86\% were obtained for EEG, peripheral signals, and the fusion scheme, respectively. And, average classification accuracies of 56.64\%, 58.64\%, and 54.79\% were obtained for EEG, peripheral signals, and the fusion scheme, respectively, in the subject-independent scenario. 
These accuracies are significantly higher than random, i.e., 50\% (one sample t-tests, $p < 0.05$), which indicates that there exists significant difference in perceptual experience between tone-mapped HDR and LDR videos. This also suggests that the perceptual experience of tone-mapped HDR videos can be determined by EEG, peripheral signals, and their combination. The best performance is obtained by the bimodal fusion and the peripheral signals for the subject-dependent and subject-independent scenarios, respectively. 

The segmentation length of physiological signals influences on the classification performance, particularly for EEG. In the subject-dependent classification scenario, average classification accuracies of 68.68\%, 69.86\%, and 58.67\% are obtained when the EEG signals are divided into 5 second-long, 10 second-long, and 20 second-long segments, respectively. Thus, the best result is obtained with 10 second-long EEG segments, which is employed in this paper.

\subsection{Classification based on questionnaire survey results}
\label{subsec:res.Q}


\begin{figure}[!t]
	\centering
	\subfloat[]{
		\includegraphics[height=1.7in]{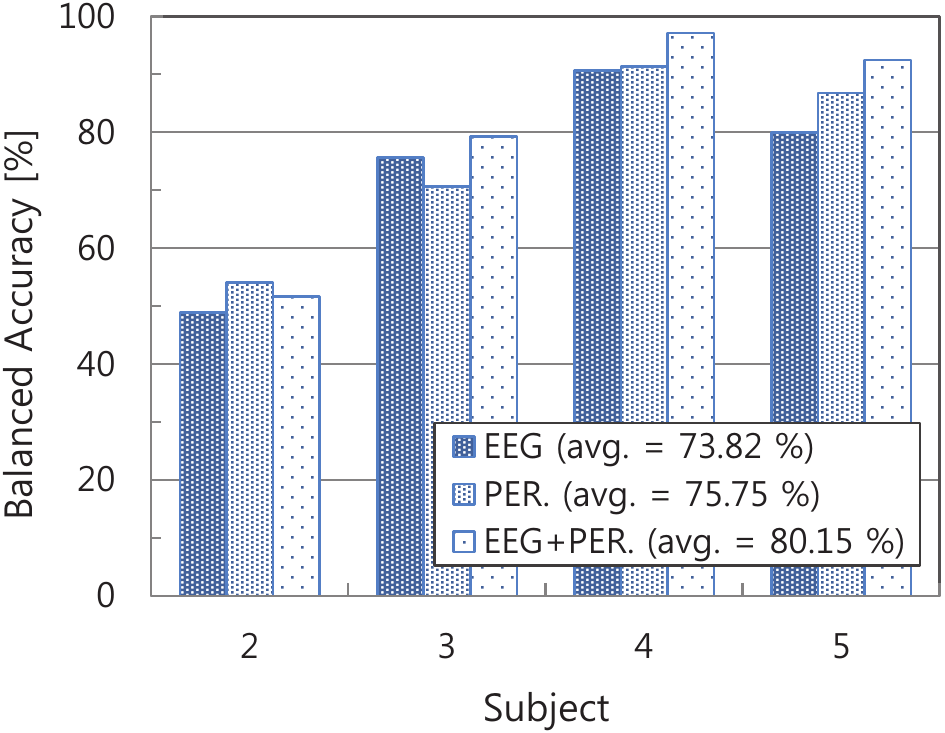}}
	
	
	\subfloat[]{
		\includegraphics[height=1.7in]{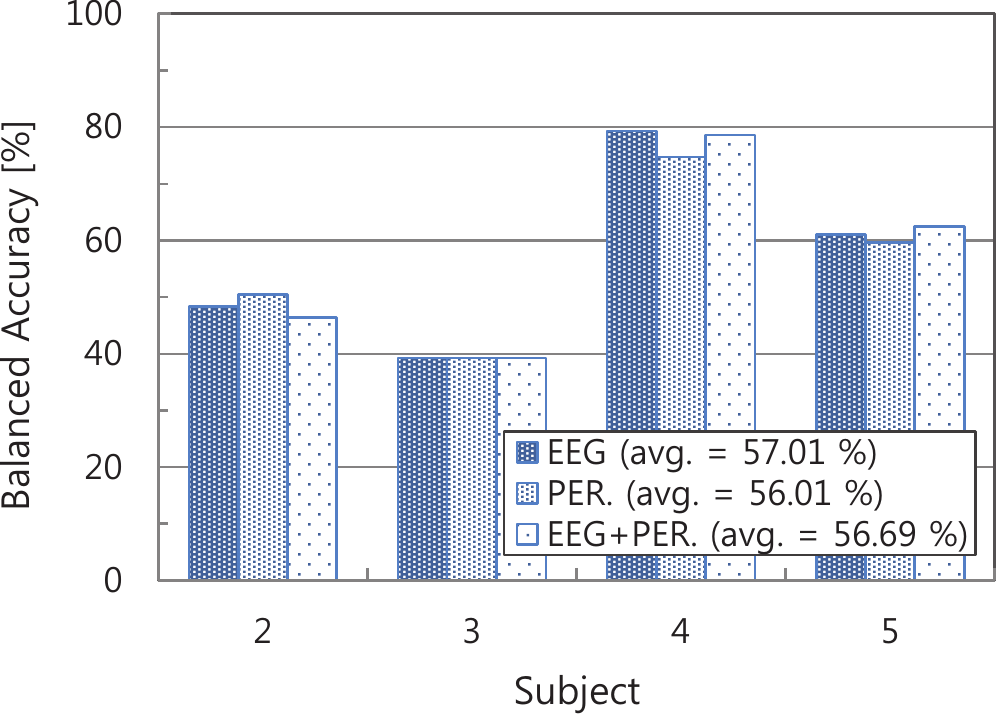}}
	
	\caption{Results of classification of contrast quality (Q1) in terms of balanced accuracy for (a) the subject-dependent scenario, and (b) the subject-independent scenario}
	\label{fig:Q1}
	
	\vspace{0.4in}
	
	\subfloat[]{
		\includegraphics[height=1.7in]{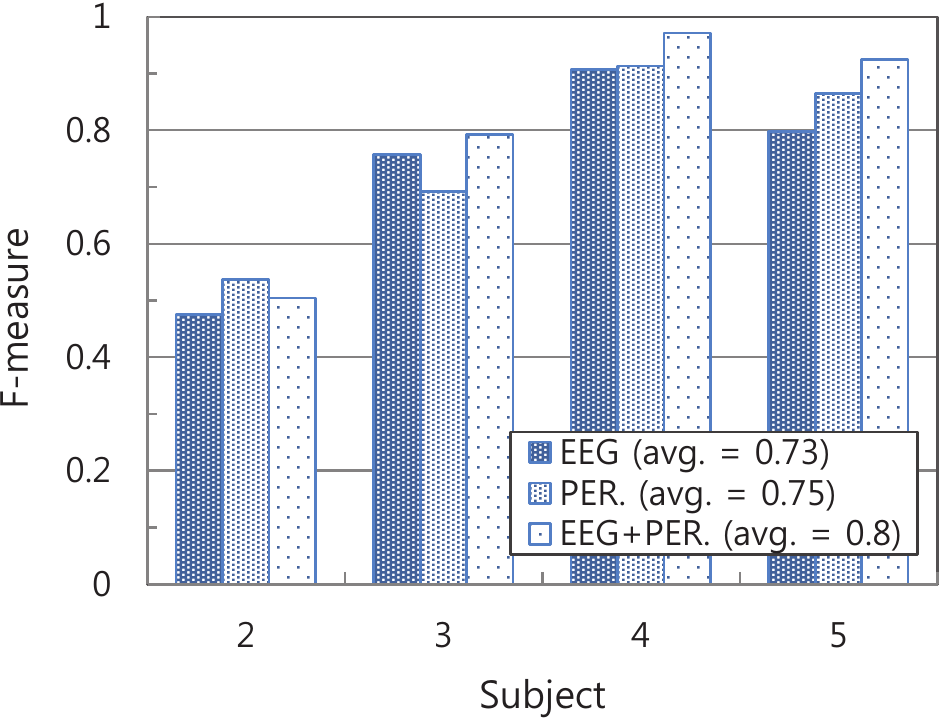}}
	
	
	\subfloat[]{
		\includegraphics[height=1.7in]{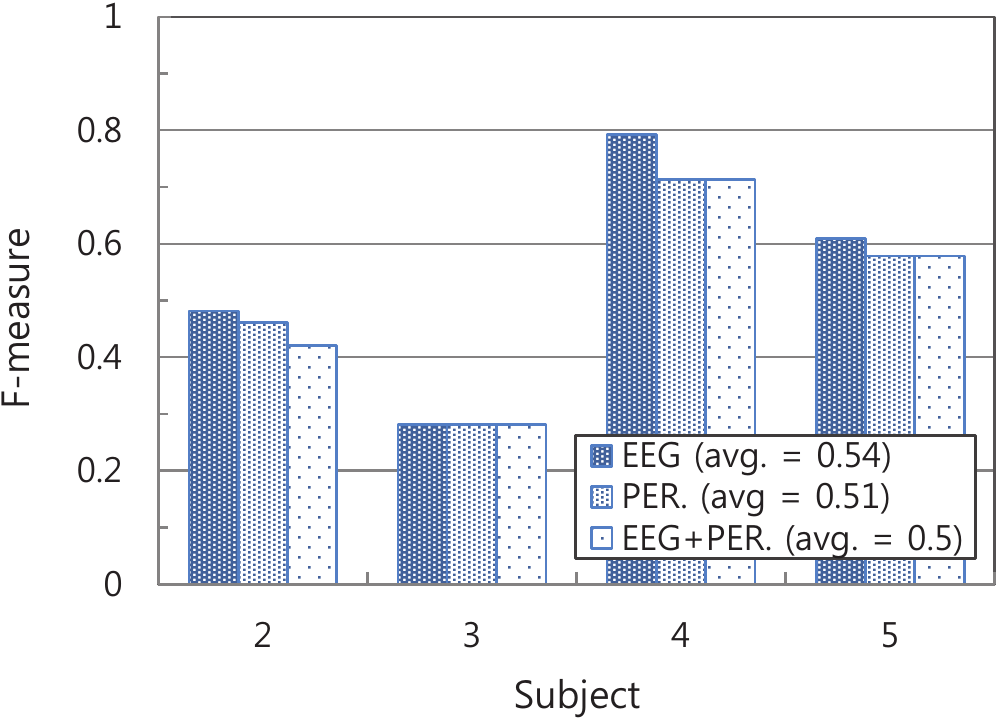}}
	\caption{Results of classification of contrast quality (Q1) in terms of F-measure for (a) the subject-dependent scenario, and (b) the subject-independent scenario}
	\label{fig:Q1F1}
\end{figure}

Figures \ref{fig:Q1} and \ref{fig:Q1F1} show the balanced classification accuracy and F-measure for classification of contrast quality, respectively. In terms of mean balanced accuracy, 73.82\%, 75.75\%, and 80.15\% were obtained  by EEG, peripheral signals, and bimodal fusion under the subject-dependent scenario. In the subject-independent scenario, mean balanced accuracies of 57.01\%, 56.01\%, and 56.59\% were obtained in the same order. One sample t-tests reveal that the results of the subject-dependent scenario are significantly higher than random ($p < 0.05$), but those of the subject-independent scenario are not ($p > 0.1$). 
In the same way, mean F-measure values of 0.73, 0.75, 0.80 and 0.54, 0.51, 0.50 are obtained for the subject-dependent and subject-independent scenarios, respectively. One sample t-tests show that only the results of the subject-dependent scenario are significantly higher than random ($p < 0.1$ for EEG and $p < 0.05$ for the others). 
The fusion scheme and EEG show the best performance in the subject-dependent and subject-independent scenarios, respectively.


\begin{figure}[!t]
	\centering
	\subfloat[]{
	\includegraphics[height=1.7in]{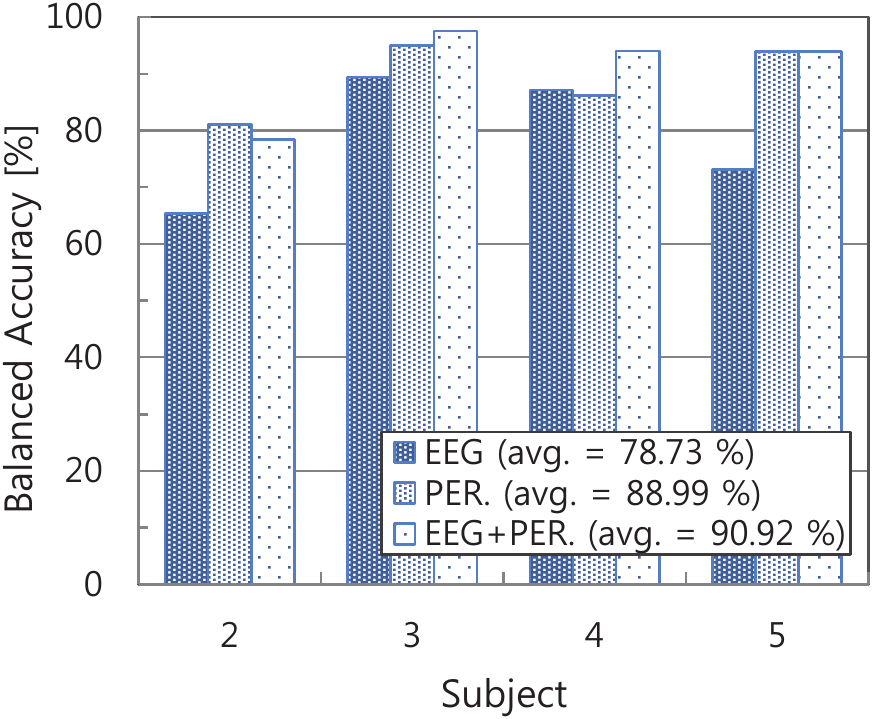}
	}
	
	
	\subfloat[]{
	\includegraphics[height=1.7in]{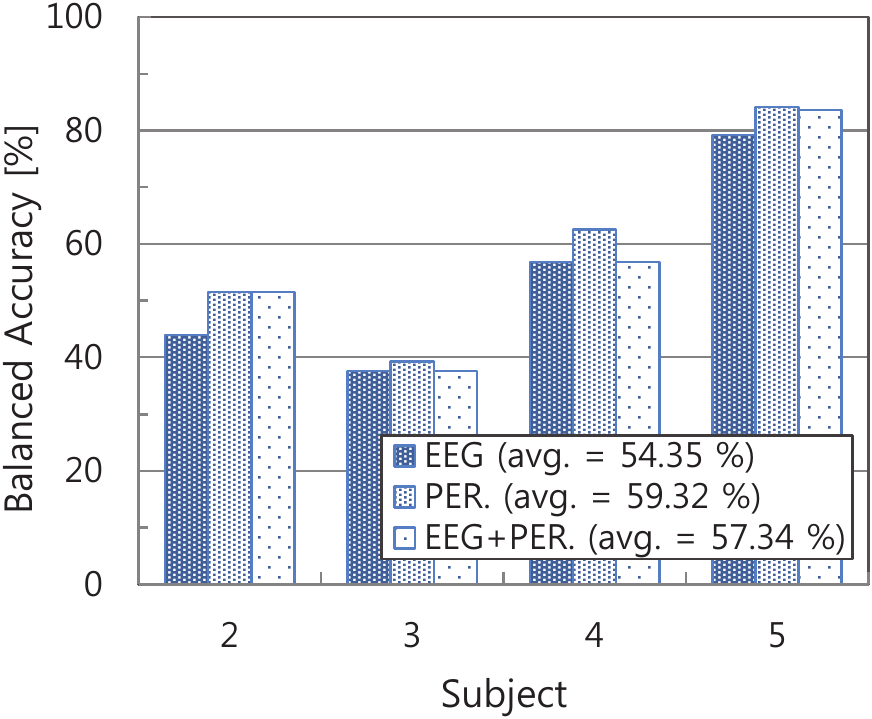}
	}
	\caption{Results of classification of overall quality (Q3) in terms of balanced accuracy for (a) the subject-dependent scenario, and (b) the subject-independent scenario}
	\label{fig:Q3}
	
	\vspace{0.4in}
	
	\subfloat[]{
	\includegraphics[height=1.7in]{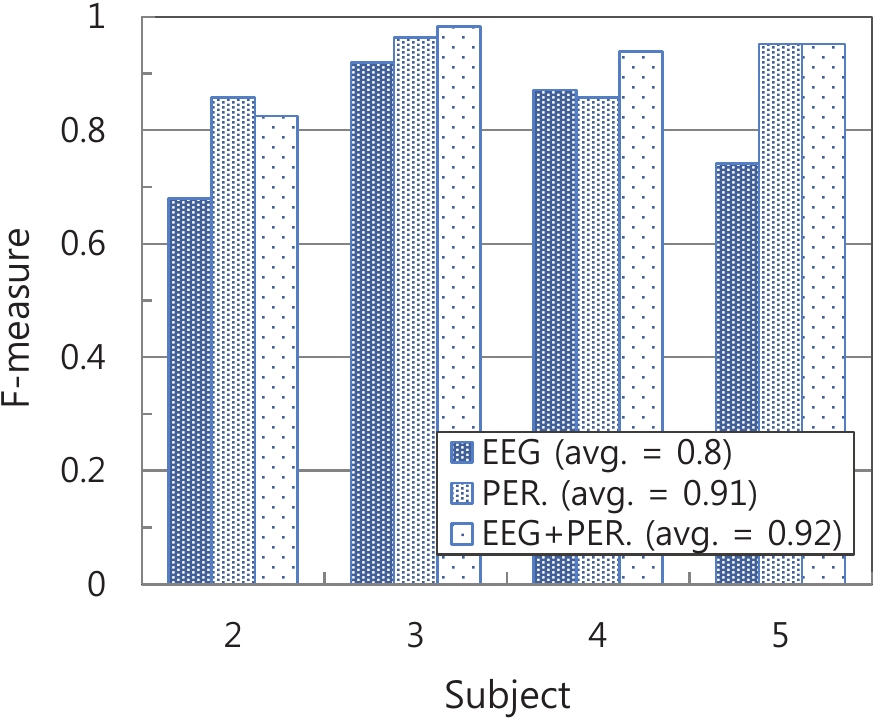}
	}
	
	
	\subfloat[]{
	\includegraphics[height=1.7in]{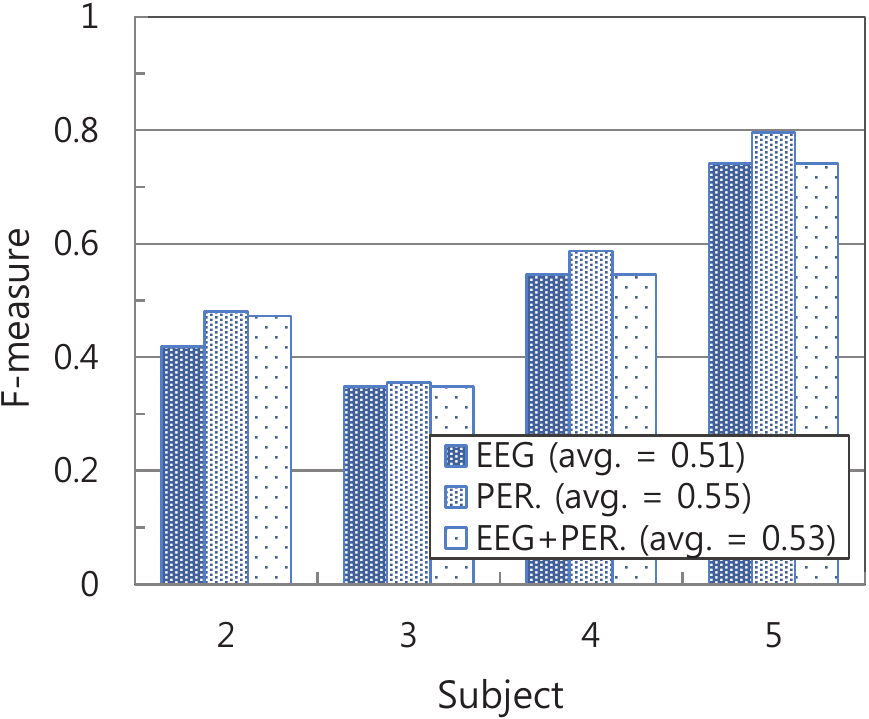}
	}
	\caption{Results of classification of overall quality (Q3) in terms of F-measure for (a) the subject-dependent scenario, and (b) the subject-independent scenario}
	\label{fig:Q3F1}
\end{figure}

The results of classification based on overall quality are shown in Figures \ref{fig:Q3} and \ref{fig:Q3F1}. Mean balanced accuracies of 78.73\%, 88.99\%, and 90.92\% were obtained from EEG, peripheral signals, and the fusion scheme, respectively, in the subject-dependent scenario. In the same order, balanced accuracies of 54.35\%, 59.32\%, and 57.34\% were obtained in the subject-independent scenario. 
Mean F-measures of 0.80, 0.91, and 0.92 were acquired from EEG, peripheral signals, and the fusion scheme, respectively. Also, in the same order, 0.51, 0.55, and 0.53 were obtained as mean F-measures in the subject-independent scenario. For both performance measures, the results of the subject-dependent scenario are significantly higher than random ($p < 0.01$), but those of the subject-independent scenario are not ($p > 0.1$).
The fusion scheme shows the best performance in the subject-dependent scenario, and the peripheral signals show the best performance in the subject-independent scenario.

\subsection{Feature analysis}
\label{subsec:res.feature}

\begin{figure}[!t]
	\centering
	\includegraphics[width=2.7in]{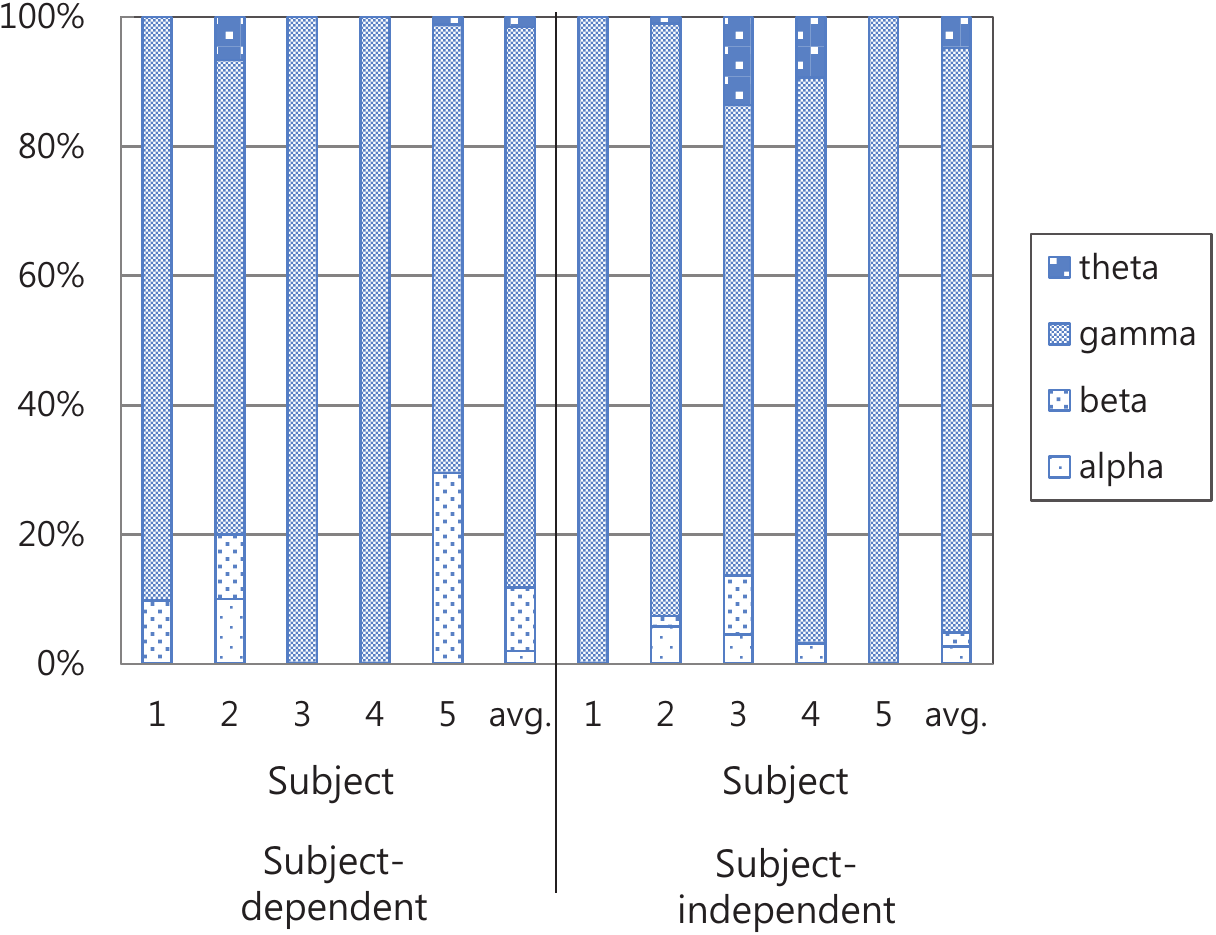}
	\caption{Relative frequencies of the selected EEG features in the classification between tone-mapped HDR and LDR videos}
	\label{fig:featureratio_tmd}
	
	\vspace{0.2in}
	
	\centering
	\subfloat[]{
	\includegraphics[width=2.7in]{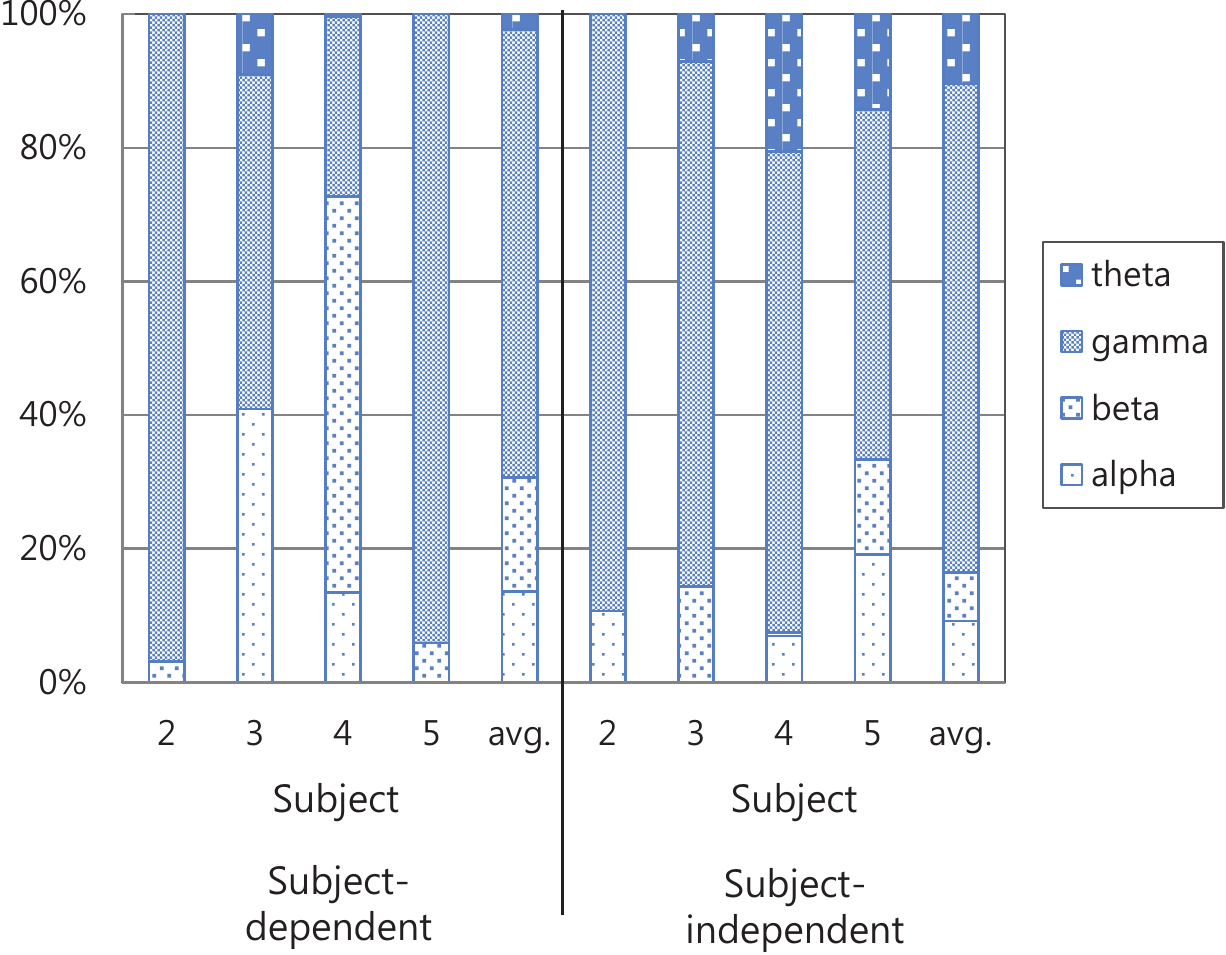}
	}
	
	
	\subfloat[]{
	\includegraphics[width=2.7in]{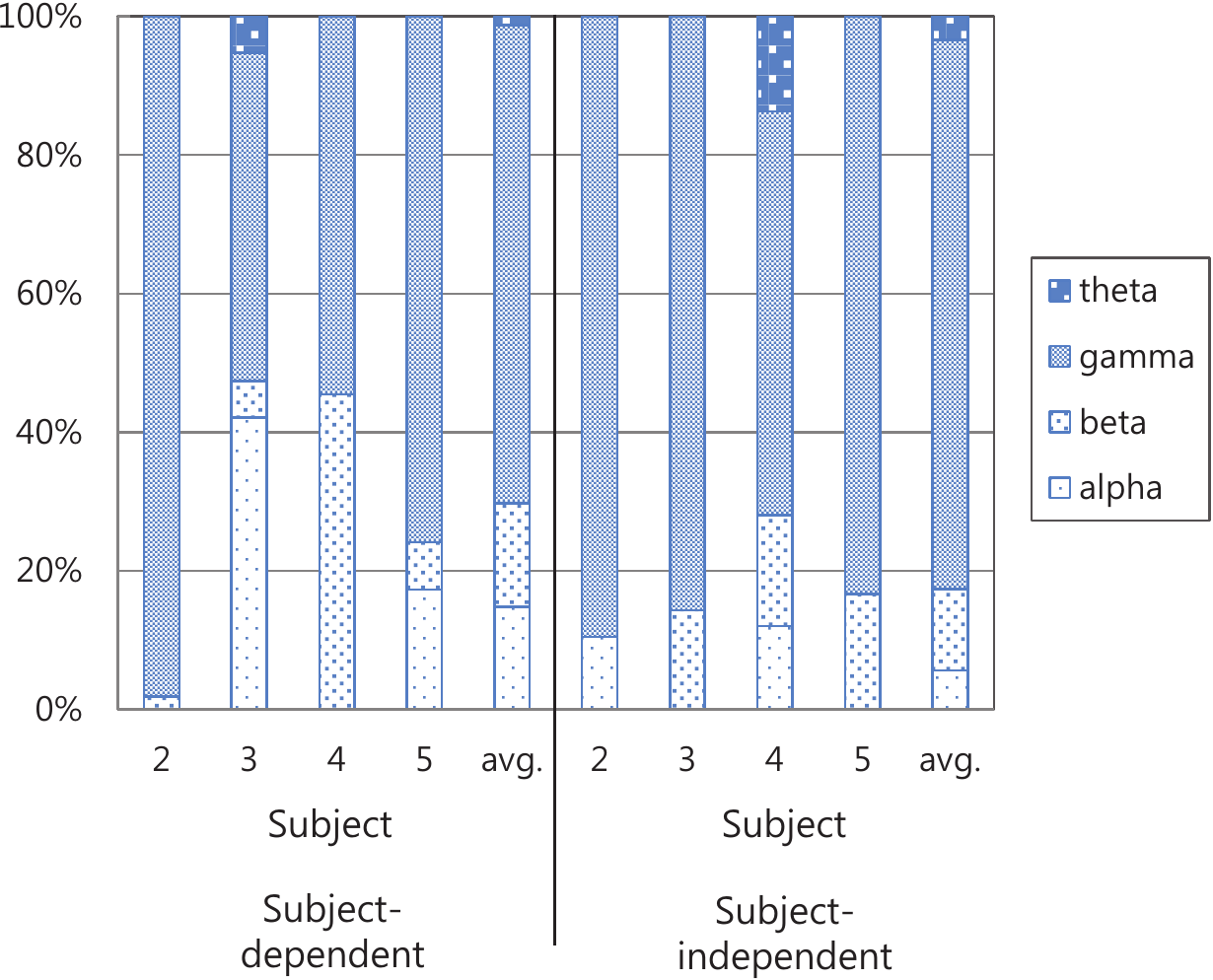}
	}
	\caption{Relative frequencies of the selected EEG features in the classification based on (a) contrast quality (Q1) (b) overall quality (Q3)}
	\label{fig:featureratio_Q}
\end{figure}

\begin{figure}[!t]
	\centering
	\includegraphics[width=2.7in]{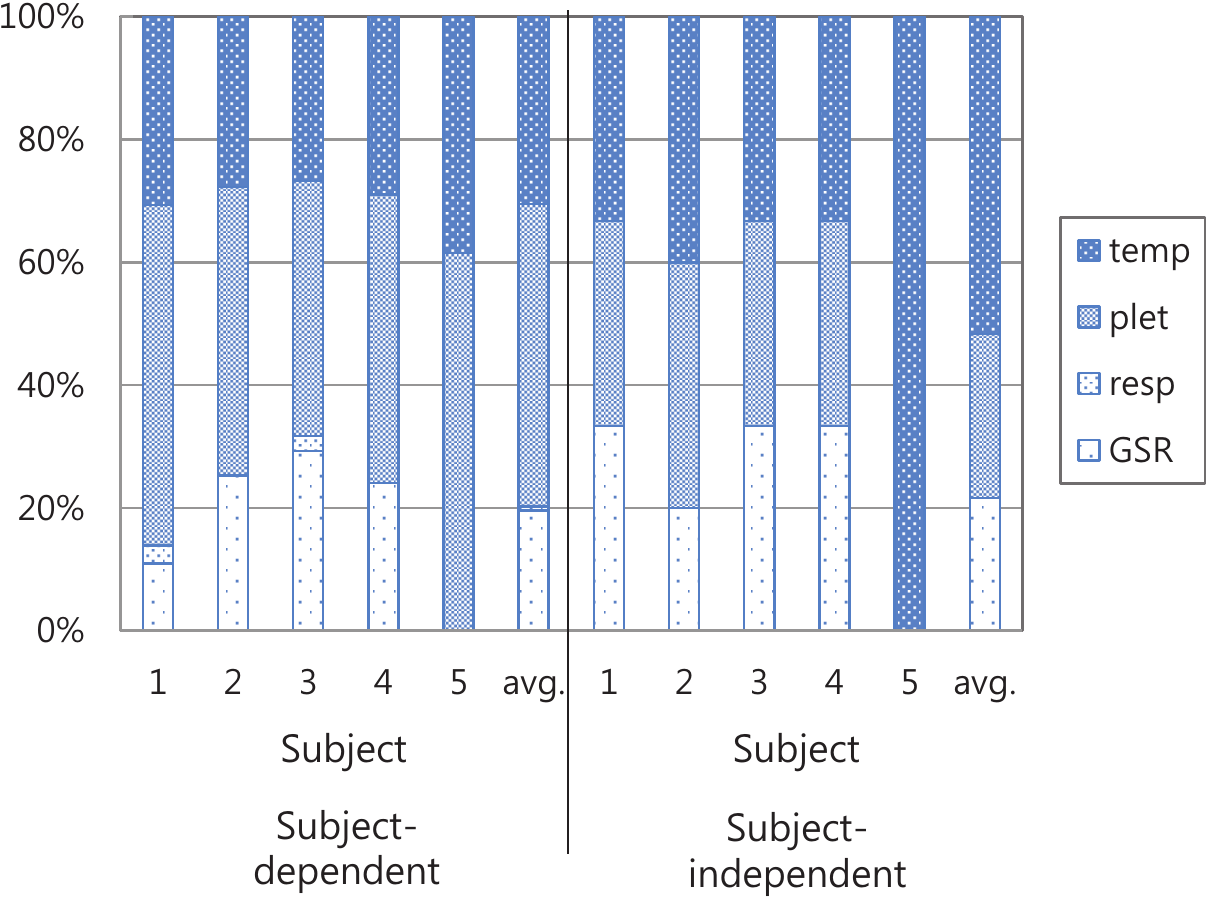}
	\caption{Relative frequencies of the selected peripheral physiological signal features in the classification between tone-mapped HDR and LDR videos}
	\label{fig:featureratio_tmd_per}
	
	\vspace{0.2in}
	
	\centering
	\subfloat[]{
	\includegraphics[width=2.7in]{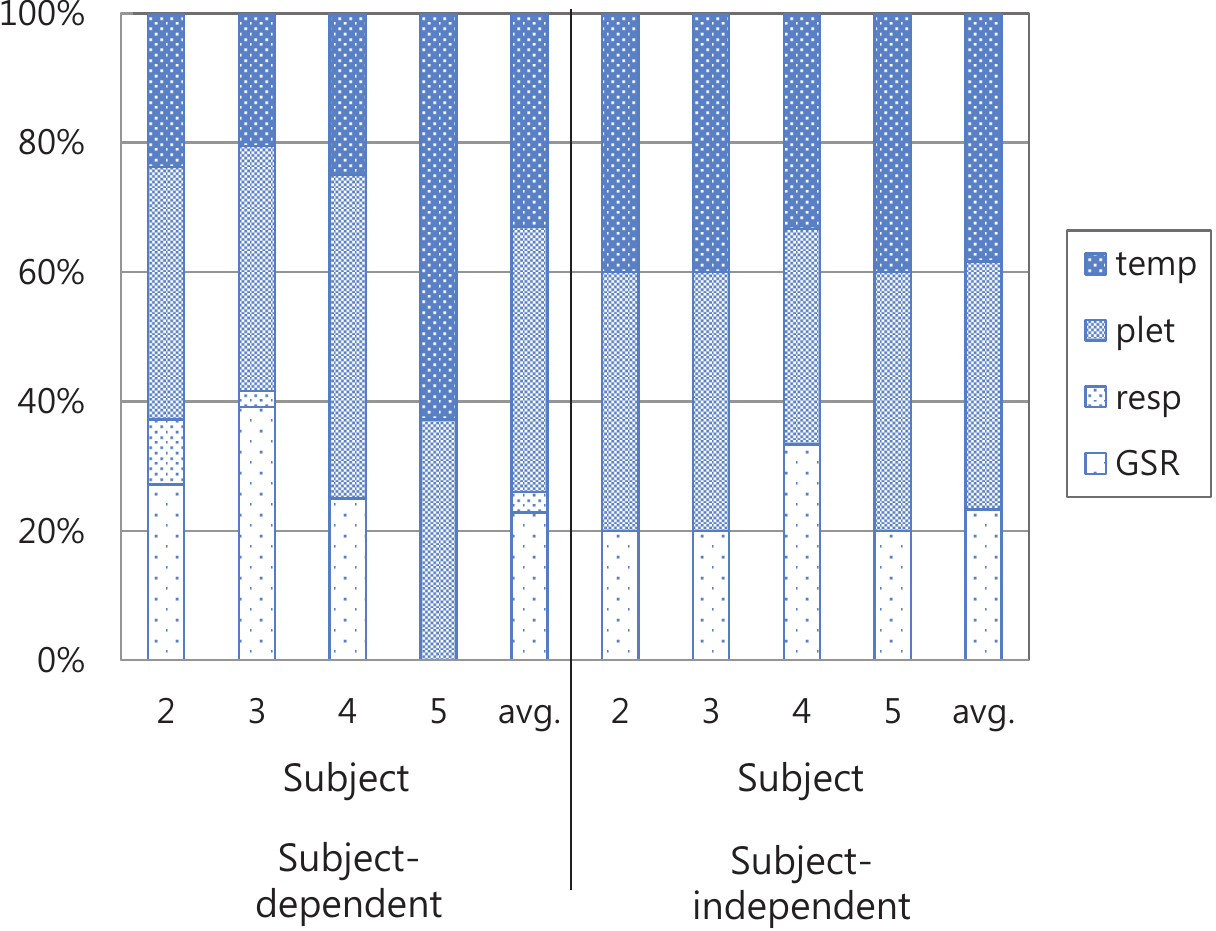}
	}
	
	
	\subfloat[]{
	\includegraphics[width=2.7in]{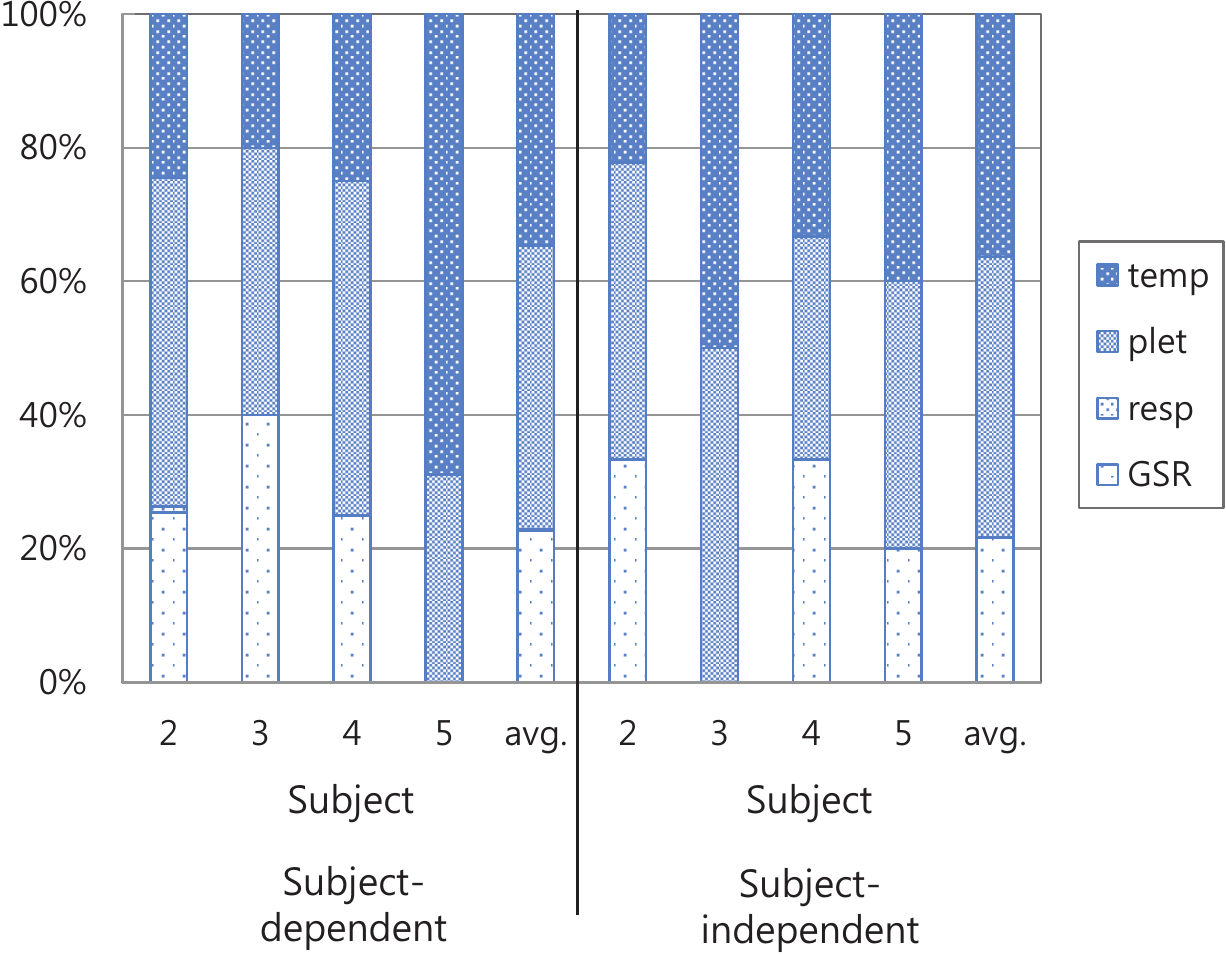}
	}
	\caption{Relative frequencies of the selected peripheral physiological signal features in the classification based on (a) contrast quality (Q1) (b) overall quality (Q3)}
	\label{fig:featureratio_Q_per}
\end{figure}

Figures \ref{fig:featureratio_tmd} and \ref{fig:featureratio_Q} present the relative frequencies of the selected EEG features with respect to frequency bands. 
It can be observed that the features extracted from the gamma band are mainly employed by the classification systems. 

\begin{figure*}[!t]
	\centering
	\includegraphics[width=6in]{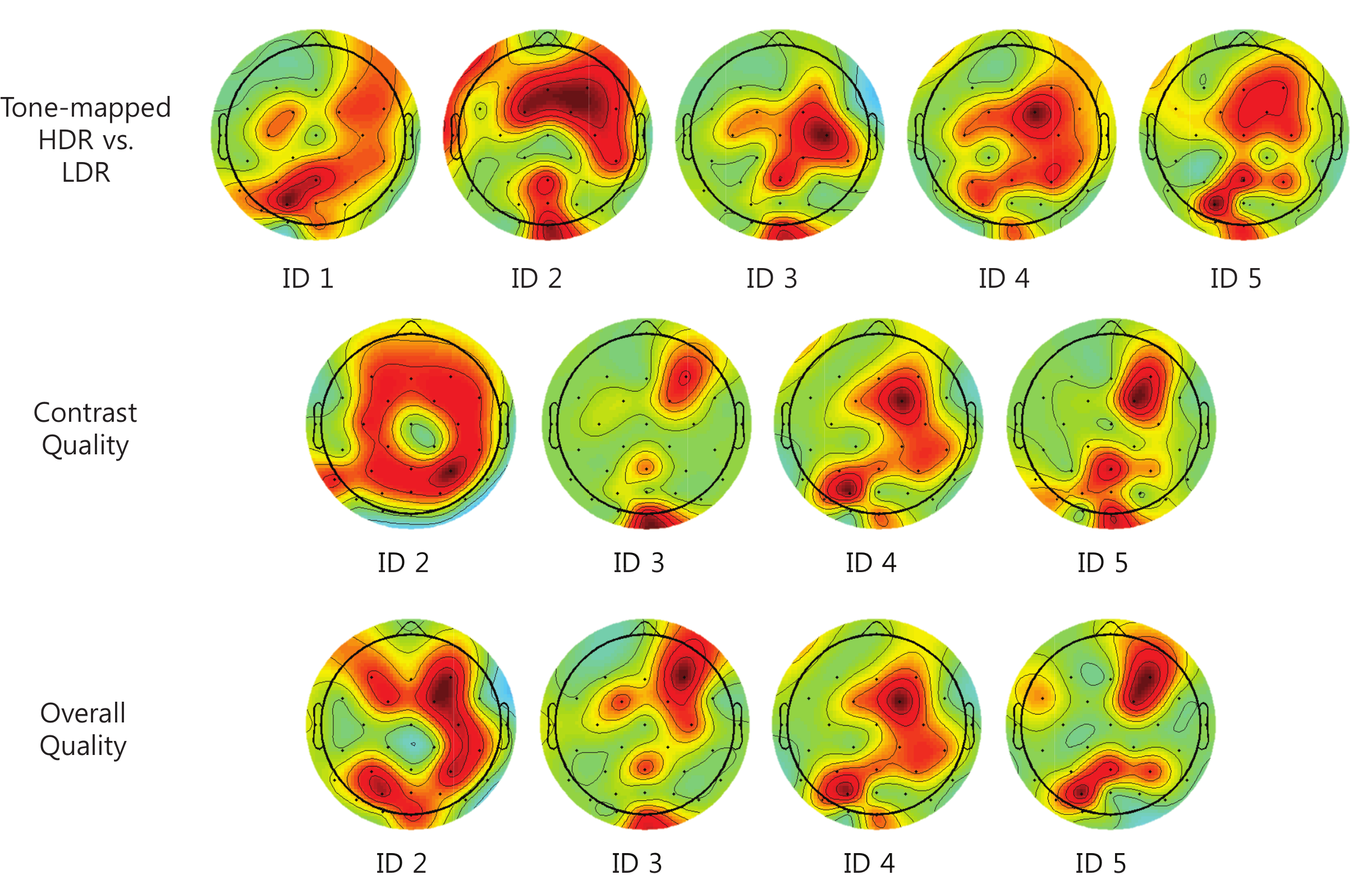}
	\caption{Significance of the EEG features chosen for the subject-dependent classification}
	\label{fig:tomo_dep}
\end{figure*}

However, we found that spatial locations of the significant features vary across the subjects. In Figure \ref{fig:tomo_dep}, the significance of EEG features is described in terms of the number of times chosen for the subject-dependent classification. In other words, the area marked by dark red color corresponds to an electrode channel employed 28 times by the classification system, and the area marked by blue color corresponds to an electrode channel that was never employed by the classification system. It is difficult to find similarity in significant feature locations across the subjects. This proves the existence of individual difference in perceptual processing of visual stimuli, which is considered as one of the causes of relatively low classification performance in the subject-independent scenario.

The relative frequencies of the selected features for the peripheral physiological signals are shown in Figures \ref{fig:featureratio_tmd_per} and \ref{fig:featureratio_Q_per}. All the peripheral channels except for respiration appear relevant for classification, with slight emphasis on the plethysmography and skin temperature.

\subsection{Discussion}
\label{subsec:dis}

Overall, the classification results shown above indicate that perceptual experience of tone-mapped HDR videos can be observed through EEG, peripheral signals, and their combination. However, the performance of the classification in the subject-independent scenario is relatively lower than that in the subject-dependent scenario, which seems to stem partially from the subject-wise discrepancy on the location of the selected features. In particular, while the performance of the classification between tone-mapped HDR and LDR videos is significantly higher than random even in the subject-independent scenario, significance of the subject-independent classification performance is not observed for classifications of contrast quality and overall quality. This also supports that quality perception of LDR and HDR videos is highly subjective. 

When the two modalities are compared, superiority of one modality against another is not consistent; classification using EEG sometimes outperforms that using peripheral signals, and sometimes vice versa. The classification performance is enhanced by the fusion scheme under the subject-dependent scenario. This proves the complementariness of EEG and peripheral signals for classification.
However, the performance is degraded via fusion under the subject-independent scenario. 

Such failure is due to inaccurate estimation of the fusion weight determining relative importance of each modality. In order to examine the reliability of the obtained fusion weight through training, the optimal weight values were exhaustively searched for the test data and compared to the trained values. The root-mean-square errors (RMSEs) between the trained and the optimal weight values were 0.0357 (0.0462 for HDR vs. LDR, 0.0377 for high vs. low contrast quality, and 0.0078 for high vs. low overall quality) for the subject-dependent scenario and 0.5031 (0.4313 for HDR vs. LDR, 0.4770 for high vs. low contrast quality, and 0.6021 for high vs. low overall quality) for the subject-independent scenario. The large RMSEs in the latter demonstrate that the relative importance of the two modalities significantly varies across subjects, which is due to individual differences in the physiological responses.

The feature analysis revealed that the gamma frequency band of EEG is significantly associated with tone-mapped HDR perception. It is known that the activity of the gamma frequency band is associated with sensory stimuli involving the auditory and visual systems \cite{Galambos:gamma}. In \cite{Tallon:gamma}, it was found that a top-down visual process enhances the oscillatory gamma response. In \cite{Muller:gamma}, the relation between the gamma band activity and human visual information processing was studied, which showed that the power of the gamma band increases when subjects pay more attention to a visual stimulus, in comparison to the case where the same stimulus is ignored. Based on these, it can be inferred that, in comparison to LDR videos, (tone-mapped) HDR videos provide viewers with more visual information via enhanced contrast and thus induce higher activities in the gamma band of EEG.

From the feature analysis of the peripheral physiological signals, on the other hand, heart rate, skin temperature, and GSR are all shown to be relevant for classification. Previous studies have demonstrated the relationship between the amount of processed information and GSR \cite{Swart76}, the relationship between mental workload and heart rate \cite{Saykrs73}, and the relationship between color preference and skin temperature \cite{Genno97}, which support our results regarding observability of distinguished peripheral physiological responses for HDR videos.

\section{Conclusion}
\label{sec:concl}

In this paper, we studied perceptual experience of tone-mapped HDR videos through explicit questionnaire survey and implicit monitoring of physiological signals.
In the explicit QoE measurement, tone-mapped HDR videos showed better perceptual experience in terms of contrast, interest, naturalness, and overall quality. In the implicit user monitoring of EEG and peripheral physiological responses via unimodal and bimodal classification, statistically significant accuracy was obtained for classification of physiological signals with respect to the dynamic range of videos in both the subject-dependent scenario and the subject-independent scenario. Significance of performance was also observed in the classification based on contrast and overall quality under the subject-dependent scenario.
Further, the EEG features chosen for classification were studied. Although the gamma band was shown to be significant for classification, difference in locations of the chosen features across the subjects was observed. This indicates existence of individual difference in perception of HDR visual stimuli, which causes difficulty in subject-independent classification of physiological signals. Also, the features extracted from skin temperature, plethysmography, and GSR were shown to be relevant, while those from respiration were not.

Upon the aforementioned valuable findings, further studies employing more subjects  would be desirable. In particular, a larger number of subjects may improve the performance of the classification system in the subject-independent scenario.

\section*{Acknowledgment}
This work was supported by the MSIP (Ministry of Science, ICT and Future Planning), Korea, under the ``IT Consilience Creative Program'' (NIPA-2014-H0201-14-1002) supervised by the NIPA (National IT Industry Promotion Agency), and by the Basic Science Research Program through the National Research Foundation of Korea funded by the MSIP (2013R1A1A1007822).

\ifCLASSOPTIONcaptionsoff
  \newpage
\fi



\bibliographystyle{IEEEtran}
\bibliography{bare_jrnl_rv.bib}
%
%
%

%

\begin{IEEEbiography}[{\includegraphics[width=1in,height=1.25in,clip,keepaspectratio]{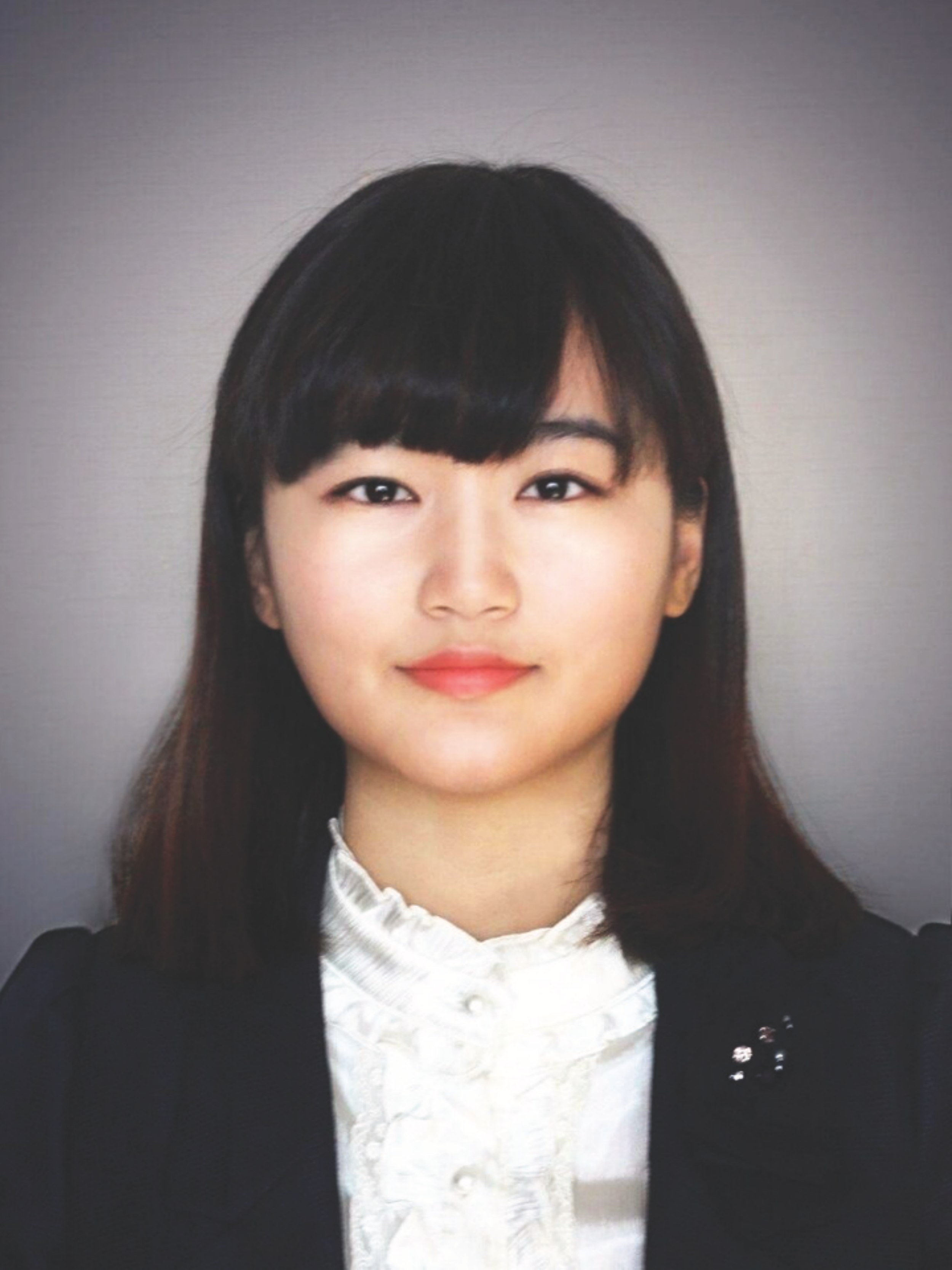}}]{Seong-Eun Moon}
 received her B.S. degree in mechanical engineering from Chiba University, Japan, in 2013. She is now with the School of Integrated Technology of Yonsei University and is working toward the Ph.D. degree. Her research interests include multimedia signal processing and physiological signal processing.
\end{IEEEbiography}

\begin{IEEEbiography}[{\includegraphics[width=1in,height=1.25in,clip,keepaspectratio]{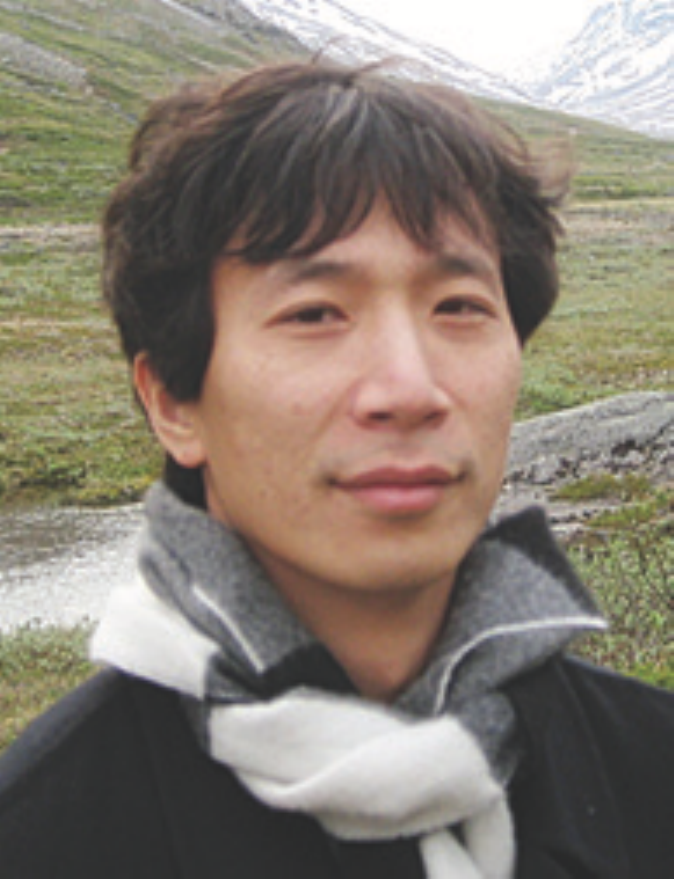}}]{Jong-Seok Lee}
 (M'06-SM'14) received his Ph.D. degree in electrical engineering and computer science in 2006 from KAIST, Korea, where he also worked as a postdoctoral researcher and an adjunct professor. From 2008 to 2011, he worked as a research scientist at Swiss Federal Institute of Technology in Lausanne (EPFL), Switzerland. Currently, he is an assistant professor in the School of Integrated Technology at Yonsei University, Korea. His research interests include multimedia signal processing and machine learning. He is an author or co-author of about 100 publications. He serves as an Area Editor for the Signal Processing: Image Communication.
\end{IEEEbiography}





\end{document}